\documentclass[twocolumn,10pt]{IEEEtran}
\DeclareUnicodeCharacter{3000}{ }

\usepackage[bottom=0.9in,top=0.7in,left=0.625in,right=0.625in]{geometry}

\ifCLASSINFOpdf
  \usepackage[pdftex]{graphicx}
  \graphicspath{{../pdf/}{../jpeg/}}
  \DeclareGraphicsExtensions{.pdf,.jpeg,.png}
\else
  \usepackage[dvips]{graphicx}
  \graphicspath{{../eps/}}
  \DeclareGraphicsExtensions{.eps}
\fi
\usepackage{enumitem}
\usepackage{epstopdf}
\usepackage{multirow}
\usepackage{float}
\usepackage[cmex10]{amsmath}
\usepackage {amssymb}
\usepackage{graphicx}
\usepackage{array}
\usepackage{mdwmath}
\usepackage{mdwtab}
\usepackage{eqparbox}

\usepackage{nomencl}
\usepackage{cite}
\usepackage{algorithm}
\usepackage{algorithmic}
\usepackage{mathtools}
\usepackage{makeidx}
\usepackage{ifthen}
\usepackage{subfigure}
\usepackage{caption}
\usepackage{color}
\usepackage{multirow,array}
\usepackage{pdflscape}
\usepackage{bm,upgreek}

\makenomenclature

\usepackage{epstopdf}
\usepackage{multirow}
\usepackage{float}
\usepackage[cmex10]{amsmath}
\usepackage {amssymb}
\usepackage{graphicx}
\usepackage{array}
\usepackage{mdwmath}
\usepackage{mdwtab}
\usepackage{eqparbox}
\usepackage{nomencl}
\usepackage{cite}
\usepackage{algorithm}
\usepackage{algorithmic}
\usepackage{makeidx}
\usepackage{ifthen}
\usepackage{subfigure}
\usepackage{caption}
\usepackage{amsthm}

\renewcommand{\nomgroup}[1]{%
\ifthenelse{\equal{#1}{I}}{\item[\textbf{Indices}]}{%
\ifthenelse{\equal{#1}{A}}{\item[\textbf{Abbreviations}]}{%
\ifthenelse{\equal{#1}{V}}{\item[\textbf{Variables}]}{%
\ifthenelse{\equal{#1}{P}}{\item[\textbf{Parameters and Constants}]}{%
}
}
}
}
}

\ifCLASSINFOpdf
  \usepackage[pdftex]{graphicx}
  \graphicspath{{../pdf/}{../jpeg/}}
  \DeclareGraphicsExtensions{.pdf,.jpeg,.png}
\else
  \usepackage[dvips]{graphicx}
  \graphicspath{{../eps/}}
  \DeclareGraphicsExtensions{}
\fi
\usepackage{epstopdf}

\usepackage{multirow,tikz,enumitem}
\usepackage{xspace,amsmath,amsfonts,amssymb,tikz,multirow,float,newclude,pgfplots, stmaryrd,lipsum,mathtools, tikz, amsthm,changepage }
\usepackage{xspace,amsmath,amsfonts,amssymb,tikz,multirow, pgfplots,pbox}
\usepackage{caption}
\usepackage{xspace}
\usepackage{latexsym}
\usepackage{xcolor}
\usepackage{graphicx}
\usepackage{enumitem}
\usepackage{titlesec}
\usepackage{comment,verbatim}
\usepackage{array}
\usepackage{booktabs}
\usepackage{titlesec}
\usepackage{cite}
\usepackage{fancyhdr}

\usepackage{comment}
\usepackage{epstopdf}
\usepackage{float}

\newcommand{\beq}{\begin{equation}}
\newcommand{\eeq}{\end{equation}}
\newcommand{\beqn}{\begin{eqnarray}}
\newcommand{\eeqn}{\end{eqnarray}}
\newcommand{\beqno}{\begin{eqnarray*}}
\newcommand{\eeqno}{\end{eqnarray*}}
\newcommand{\bma}{\begin{displaymath}}
\newcommand{\ema}{\end{displaymath}}
\newcommand{\bnu}{\begin{enumerate}}
\newcommand{\enu}{\end{enumerate}}
\newcommand{\bce}{\begin{center}}
\newcommand{\ece}{\end{center}}
\newcommand{\btb}{\begin{tabular}}
\newcommand{\etb}{\end{tabular}}

\def\bx{{\mathbf{x}}}

\def\bz{{\mathbf{z}}}
\def\bv{{\mathbf{v}}}

\def\bu{{\mathbf{u}}}

\def\by{{\mathbf{y}}}
\def\one{{\mathbf{1}}}

\def\a{\alpha}
\def\g{\gamma}
\def\re{\mathbb{R}}
\def\T{\mathsf{T}}
\def\E{\mathsf{E}}
\def\G{\mathbb{G}}

\def\bF{{\mathbf{F}}}

\def\cNik{{\mathcal{N}^{\rm }_{ik}}}
\def\cNjk{{\mathcal{N}^{\rm }_{jk}}}

\def\la{{\langle}}
\def\ra{{\rangle}}

\newtheorem{theorem}{Theorem}
\newtheorem{lemma}{Lemma}
\newtheorem{remark}{Remark}

\newtheorem{assumption}{Assumption}

\newtheorem{definition}{Definition}

\allowdisplaybreaks[2]

\begin{document}


\title{CrowdCache: A  Decentralized Game-Theoretic Framework for Mobile Edge Content Sharing}



\author{
    \IEEEauthorblockN{Duong~Thuy~Anh~Nguyen\IEEEauthorrefmark{2}, Jiaming Cheng\IEEEauthorrefmark{3},  Duong Tung Nguyen\IEEEauthorrefmark{2}, Angelia Nedi\'{c}\IEEEauthorrefmark{2}} \\
    \IEEEauthorblockA{\IEEEauthorrefmark{2}Arizona State University, Tempe, AZ 85281, USA,
    \{dtnguy52, duongnt, Angelia.Nedich\}@asu.edu} \\
    \IEEEauthorblockA{\IEEEauthorrefmark{3}University of British Columbia, Vancouver, BC V6T1Z4, Canada,
    jiaming@ece.ubc.ca} \vspace{-0.57em}
\thanks{The first two authors have contributed equally to this work.} 
}

\maketitle

\begin{abstract}
Mobile edge computing (MEC) is a promising solution for enhancing the user experience, minimizing content delivery expenses, and reducing backhaul traffic. In this paper, we propose a novel privacy-preserving decentralized game-theoretic framework for resource crowdsourcing in  MEC. Our framework models the interactions between a content provider (CP) and multiple mobile edge device users (MEDs)  as a non-cooperative game, in which MEDs offer idle storage resources for content caching in exchange for rewards. We introduce efficient decentralized gradient play algorithms for Nash equilibrium (NE) computation by exchanging local information among neighboring MEDs only, thus preventing attackers from learning users' private information. The key challenge in designing such algorithms is that  communication among MEDs is not fixed and is facilitated by a sequence of undirected time-varying graphs. Our approach achieves linear convergence to the NE without imposing any assumptions on the values of parameters in the local objective functions, such as requiring strong monotonicity to be stronger than its dependence on other MEDs' actions, which is commonly required in existing literature when the graph is directed time-varying. Extensive simulations demonstrate the effectiveness of our approach in achieving efficient resource outsourcing decisions while preserving the privacy of the edge devices. 
\end{abstract}


\printnomenclature

\section{Introduction}
\label{intro}


The rapid proliferation of data-intensive mobile devices and the growing demand for high-quality services have resulted in an unprecedented surge in mobile data traffic.
This surge in traffic  has placed a significant burden on mobile networks, which have been identified as the bottleneck of the mobile Internet. 
Compounding this issue, multiple mobile edge devices (MEDs) repeatedly downloading the same popular content can result in a significant waste of backhaul resources and severe network congestion, which ultimately impacts the quality of service (QoS).
Hence, there is an urgent need for innovative solutions to alleviate the strain on the network.

Centralized systems typically store content on remote cloud servers or content delivery networks (CDNs). When a user requests content, the request is transmitted to a nearby base station, which forwards it through the backhaul network to the core network. The core network then retrieves the content from the centralized servers or CDNs and sends it back through the backhaul network to the base station for delivery to the user's device. However, this reliance on backhaul networks can cause network congestion and slow delivery times, particularly during high-traffic periods. Edge caching has emerged as a promising technology that can help alleviate these challenges.

The reduction in data storage cost has paved the way for edge caching, a technology that proactively caches popular content at the network edge 
during off-peak hours. As a result, requests during peak hours can be served locally at the edge, rather than traversing the mobile core and the internet to reach the original servers. This approach can substantially  reduce duplicate data transmission, decrease backhaul capacity requirements, alleviate congestion within the backbone network, and ultimately improve the overall user experience \cite{yang2012,jiang20}. 
In addition to content caching at MEC servers situated at base stations or edge data centers, the idle resources of numerous MEDs can be utilized for edge caching to  reduce the upfront investment costs for content providers (CPs) and network operators.
Implementing mobile edge caching requires active participation from many MEDs who are willing to contribute their storage resources. To address this challenge, we propose CrowdCache, a novel crowdsourcing-based mobile edge caching and sharing service that distributes the caching workload and reduces reliance on centralized caching infrastructure. 
\vspace{-0.3cm} 
\begin{figure}[ht!]
	\centering
		\includegraphics[width=0.45\textwidth,height=0.11\textheight]{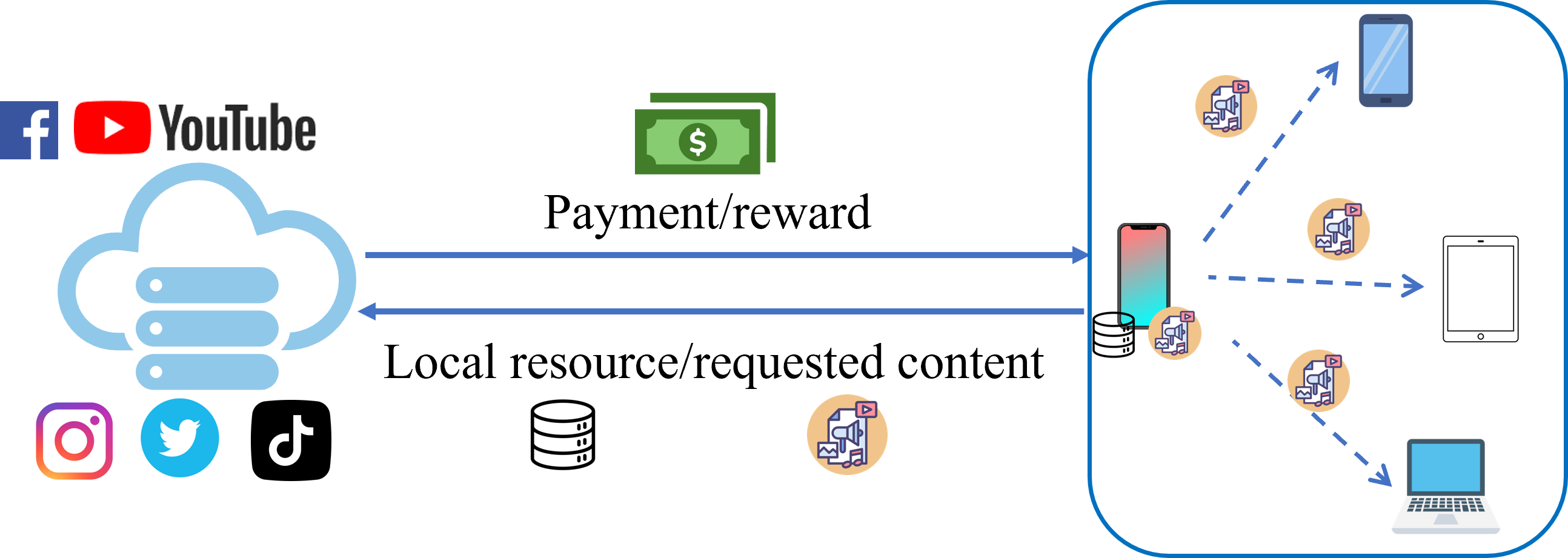}
			\caption{CrowdCache framework}
	\label{fig:CrowdCache}
\end{figure} \vspace{-0.3cm} 

Figure \ref{fig:CrowdCache}
depicts the architecture of CrowdCache, which is a decentralized caching system that utilizes idle resources from MEDs to cache popular content at the network edge.
The system consists of a CP and multiple MEDs. 
By crowdsourcing idle resources from MEDs, the CP can virtually expand the total edge caching capacity and drastically improve user experience. 
 In a given service area, popular content such as highly-rated movies, viral videos, news updates, and weather forecasts are often requested by numerous users.  Therefore, rather than accessing remote content servers, mobile users can retrieve a portion of content directly from nearby MEDs that participate in the CrowdCache system. 
Specifically, the CP can store and deliver cached content on these participating MEDs to fulfill requests from nearby users. We consider a CP that provides public content. Thus, potential data privacy concerns related to the cached content are not significant factors to consider.

The economic incentives for MEDs to participate in content sharing  are pivotal to the success of the CrowdCache system.  MEDs and the CP are self-interested parties motivated to maximize their respective benefits. 
Proactive caching at the edge can reduce operational expenses and improve service quality for the CP.
The CP can offer various incentives, including monetary compensation, reduced data plan costs, or free access to premium content to the contributing MEDs. 
However, there is an inherent tradeoff between the reward scheme and the benefits gained by caching content on MEDs. 
MEDs can monetize their idle resources by contributing to the CrowdCache system. MEDs may decide to participate in the CrowdCache system based on the incentives offered, weighing the associated costs, such as increased power consumption and reduced battery life. To optimize their utilities,  MEDs must determine the optimal amount of storage to contribute to CrowdCache.  
Ultimately, CrowdCache enhances the efficiency and effectiveness of available resources, resulting in an improved experience for all users.

In designing a decentralized system, considerations must go beyond economics and incentive mechanism design as there are other important aspects that need to be addressed. One of the major concerns is \textit{privacy}, which can be compromised in centralized algorithms if the system lacks strong privacy protections such as data encryption, differential privacy, or secure multi-party computation. In our setting, MEDs may not want to reveal their utility functions and physical constraints (e.g., caching capacity) to the system since the system may exploit this information to minimize the reward to the MEDs. Even in a distributed system,  the centralized coordinator (i.e., the CP in our setting) collects data from all MEDs and rebroadcasts it at every iteration, potentially allowing them to learn patterns in the data and infer sensitive information that individual MEDs may not want to reveal. 
Furthermore, some existing frameworks require each MED to possess complete information regarding the decisions of its rivals \cite{Yu2017,Yi2019}. However, this may not be practical due to privacy concerns. For instance, the existing body of work that utilizes the best-response approach \cite{yang2012} necessitates knowledge of the local cost function, which contains sensitive information about an MED's state, resources, constraints, and preferences. Sharing this information can also reveal information about an MED's strategies or decision-making processes, which may be confidential or proprietary. Given these challenges, there is a need to develop a decentralized algorithm that relies on local information and limited information exchange to ensure privacy and confidentiality.


Extensive research has been conducted on the development of efficient distributed techniques for computing Nash-equilibrium (NE) with partial information, mainly based on projected gradient and consensus dynamics approaches. However, the early works \cite{lbo21,Yao20} have predominantly employed static communication networks, which are impractical in MEC environments. Due to changing network conditions, user mobility, 
dynamic strategies are needed to ensure efficient and effective content delivery in the presence of time-varying network among MEDs. Recent studies have addressed the NE computation in communication networks with switching topologies in \cite{nguyen2022distributed} and references therein. A key challenge is the time-varying nature of mixing matrices, \cite{nguyen2022distributed} establishes explicit bounds for the step-size and provides a condition for handling the loss of monotonicity with weighted norms. However,\textit{ this condition requires the strong monotonicity of the local objective of each MED to be sufficiently strong relative to its dependence on other MED' actions}, as noted in the literature for directed communication graphs. \textit{This limitation hinders the algorithm's direct application in the resource crowdsourcing problem under consideration}, as the local objective function's coefficient values may not always satisfy the additional assumption.

To address this limitation, this paper proposes a novel decentralized NE-seeking algorithm for a mobile edge content-sharing problem where MEDs can exchange information when located within each other's coverage areas. The communication is two-way, even though the set of devices with which each MED can communicate may change over time. \textit{The bidirectional nature of communication in our proposed MEC environment enables us to consider an undirected and time-varying communication network}, which is a departure from the directed and time-varying communication graphs considered in \cite{nguyen2022distributed}. This key difference, together with the use of Metropolis weights, results in doubly-stochastic mixing matrices that enable convergence analysis without the strong monotonicity assumption under weighted norms, overcoming the technical limitations of previous methods. Notably, the proposed algorithm achieves linear convergence without additional assumptions on the local objective function. The proposed algorithm offers a significant contribution to the MEC environment by addressing a crucial issue of dynamic network topology and high mobility MEDs. 

\textbf{Contributions.} We are among the first to examine the privacy-preserving decentralized mobile edge content caching and sharing problem, formulated through non-cooperative games while considering the high mobility of MEDs. NE offers a stable resource allocation where no MED has any incentive to change their strategy unilaterally, given the strategies of the other MEDs. We propose fully decentralized algorithms called \textbf{DCrowdCache} and \textbf{DCrowdCache-m} that utilize local information exchange among MEDs to compute the NE while preserving the privacy of MEDs. The algorithm converges over time-varying undirected communication networks by having each MED perform a gradient step to maximize their utility while sharing information with local network neighbors.  Significantly, our approach within the MEC environment achieves linear convergence to the NE without the imposition of any assumptions on the local objective function. This stands out from existing literature where the requirement for strong monotonicity to be greater than its dependence on other MEDs' actions is commonly enforced if the communication graph is  one-way and time-varying, to guarantee the strong monotonicity of the game mapping under a weighted norm. This framework has the potential to transform resource crowdsourcing in MEC environments and benefit resource-constrained MEDs while reducing energy consumption and carbon footprint. 


\section{System Model}
\label{system}
We consider a  framework for mobile edge content caching and sharing via crowdsourcing, known as \textbf{CrowdCache}. This framework comprises a cloud-based CP and a set of MEDs, which may include mobile handheld devices (e.g., smartphones) serving end-users. The CP is responsible for providing requested content to MEDs via a remote cloud server, while MEDs may request content from the CP and cache some of it on their device storage. However, utilizing a cloud server for content delivery may lead to network congestion and slow content delivery especially during peak hours. 
To enhance system efficiency, the CP may incentivize MEDs to share content with each other in a decentralized manner. 

\textbf{Mobile edge devices (MEDs):} Consider a set $\mathcal{I}$ of $N$ MEDs interested in content sharing, where each MED is indexed by $i$. In this CrowdCache system, the hosting MED responsible for content sharing also provides storage for cached content. To facilitate resource allocation decisions, we introduce the variable $x_i \geq 0$ to denote the storage a MED is willing to provide. These decisions form a profile vector $x = [x_1,x_2,\dots,x_N]$. Each MED aims to maximize its own utility by determining its resource allocation decision, taking into account costs and rewards. Caching and sharing content incurs a  cost, which varies based on device type and depreciation. We model this cost, similar to \cite{electric1,electric3}, for MED $i \in \mathcal{I}$, as follows:
\begin{align}\label{eq:cost_function}
    c_i(x_{i}) = Q_ix_{i}^2 + h_i x_{i}, ~~ \forall i.
\end{align}
$Q_i$ is the quadratic cost, and $h_i$ is the linear cost of MED $i$. This quadratic cost function
discourages excessive storage allocation. When an MED chooses to participate in the CrowdCache system, it must reserve a specific storage space for caching content, which results in a storage cost. The MED is also responsible for content delivery, which can result in significant bandwidth and power consumption, potentially leading to reduced performance and battery life for the contributing MEDs.


In addition to considering the benefits and costs associated with content caching and sharing, mobile users must also take into account physical limitations, such as the storage capacity of their devices. 
Denoting the maximum storage capacity available for cached content on MED $i$ as $C_i$, we require:
\begin{align}
\label{constr:res_cap}
    0 \leq x_i \leq C_i, ~~ \forall i.
\end{align}

\textbf{Content provider (CP):} Within the CrowdCache framework, the CP can leverage edge caching to provide the most current content to users. When a user requests content, the request is initially directed to the closest MED that has the requested cached content stored. If the MED has the content cached, it will respond to the requesting user with the content. The CP can reduce its content delivery cost significantly by encouraging content sharing among the MEDs. In return for participating in content sharing, the platform provides a reward (i.e., payment) to each MED that shares its cached content. The CP's objective is to provide content to users while improving QoS.
In this study, the pricing strategy adopted by the CP is that of a price-maker, allowing for the adjustment of the reward for MEDs based on market demand. To incentivize MEDs $i \in \mathcal{I}$ to share cached content with others, the CP may offer higher rewards. Conversely, the CP may reduce the unit price to ensure efficient resource utilization in cases where the total outsourced resources are high. The unit price can be adjusted based on the total resources outsourced by all MEDs, thus maintaining a balance between the platform's revenue and the incentives for the MEDs. The price for each unit of resource is captured by:
\begin{align}\label{eq:unit_reward_function}
    p(x)=\bar{P}-\gamma \sum_{j \in \mathcal{I}} x_j,
\end{align}
where $\bar{P}$ denotes the maximum unit reward offered by the CP (i.e., payment to the MEDs by sharing the cached content), which can be obtained according to the historical data. Here, $\gamma$ is a weighting parameter, controlled by the CP, that measures the negative effect of the other MED's benefits by sharing content. The CP can adjust the reward through parameter $\gamma$, reflecting their attitude towards the market demand for shared content. A lower value of $\gamma$ can encourage more MEDs to outsource their idle resources. Conversely, when more resources are available, unit prices may decrease by increasing $\gamma$ as demand can be sufficiently fulfilled by existing participants. The reward for MED $i$ provided by the CP can be given by
\begin{align}\label{eq:reward_function}
    R_i(x_i,x_{-i}) = p(x)x_i, ~~ \forall i.
\end{align}
Overall, the reward function is a function of $x_i$ as well as other MEDs (i.e., $x_{-i}$) in the set $\mathcal{I}$, which penalizes MEDs for increasing the sum of $x_i$ of all MEDs $i \in \mathcal{I}$. This reward function emphasizes the marginal benefit that the reward of user $i$ is diminishing as the others intensify the engagement on content sharing. Specifically, if MED $i$ decides to be completely inactive by setting $x_i = 0$, then she will receive no reward in (\ref{eq:reward_function}). On the other hand, MED $i$ will receive a positive intrinsic reward even if all other MEDs are inactive.

Subsequently, we proceed to formulate the utility function that each MED $i$ strives to maximize, which can be represented as the discrepancy between the received reward and the associated computing cost, as follows:
\begin{align}\label{eq:UtilityFunction}
    U_i(x_{i},x_{-i}) = P_i(x_i,x_{-i}) - c_i(x_i), ~~ \forall i. 
\end{align}

\section{Game Theoretic Model and NE Analysis}
\label{analysis}
Game theory offers a systematic approach to comprehending decision-making in scenarios where multiple decision-makers aim to optimize their individual, but interconnected, objectives. In this section, we adopt a game-theoretic framework to model the crowdsourcing mobile edge content caching and sharing problem as a non-cooperative game, and we utilize the concept of NE to characterize the corresponding solution. We investigate the existence and uniqueness of the NE. The uniqueness of the NE strategy profile enables the platform to accurately predict user behavior in the system, facilitating the selection of the optimal values of $\bar{P}$ and $\gamma$ for optimal profit. 

Consider the non-cooperative game between $N$ MEDs where each MED seeks to minimize their local objective function, without coordinating with other MEDs. Denote the game by $\Gamma=(\mathcal{I},\{J_i\},\{X_i\})$ where $X_i=\left[0,C_i\right]\subseteq \mathbb{R}$ represents the action set and $J_i(\cdot)$ represents the local objective function, for each MED $i\in\mathcal{I}$, given as follows:
\begin{align}
    J_i(x_{i},x_{-i}) &= -U_i(x_{i},x_{-i}) \nonumber\\
    &=Q_ix_{i}^2 + h_i x_{i} -\Bigg(\bar{P}-\gamma \sum_{j \in \mathcal{I}} x_j\Bigg)x_i, ~~ \forall i.\label{eq:localObj}
\end{align}
Each function $J_i(x_i,x_{-i})$ depends on $x_i$ and $x_{-i}$, where $x_i\in X_i$ is the action of the MED $i$, indicating the amount of idle resources they will offer, and $x_{-i} \!\in\! X_{-i}\!=\!X_1\times\cdots\times X_{i-1}\times X_{i+1}\!\times\!\cdots\!\times\! X_N$ denotes the joint action of all MEDs except MED $i$. Denote by $x\!=\!(x_1,\ldots,x_N)$ the strategy profile consisting of all MEDs’ actions and by $X = X_1\times\cdots\times X_N$ the MEDs' joint action set. We observe the following:

\begin{lemma}\label{lem-convexGame}
The game $\Gamma$ is convex. Namely, for all $i\in\mathcal{I}$, the set $X_i$ is nonempty, convex and closed, the function $J_i(x_{i},x_{-i})$ is convex in $x_i$ over $X_i$ and continuously differentiable in $x_i$ for each fixed $x_{-i}$. Moreover, the partial derivatives of $J_i$ with respect to $x_i$, $J_{x_i}(x_{i},x_{-i})$, are continuous functions in $x$.
\end{lemma}

In order to identify stable and desirable solutions for the non-cooperative game $\Gamma$ under consideration, we employ the concept of NE which refers to a joint action where no MED has an incentive to change their strategy unilaterally. The stable NE strategy set for the game $\Gamma$ is defined as follows:
\begin{definition}
A solution to the game $\Gamma$ is a NE $x^*\in X=X_1\times\cdots\times X_N$ if for every MED $i\in\mathcal{I}$, we have:
\beqn
J_i(x_i^*,x_{-i}^*)\le J_i(x_i,x_{-i}^*), \quad \forall x_i\in X_i,
\eeqn
where $\nabla_i J_i(x_i,x_{-i})=\nabla_{x_i} J_i(x_i,x_{-i})$ for all $i\in\mathcal{I}$.
\end{definition}

For any MED $i$, Lemma~\ref{lem-convexGame} states that the action set $X_i$ is closed and convex, and the cost function $J_i(x_i,x_{-i})$ is convex and differentiable in $x_i$ for every $x_{-i}\in X_{-i}$. Given these conditions, a NE $x^*\in X$ of the game can  alternatively be characterized using the first-order optimality conditions. That is, $x^*\in X$ is a NE of the game if and only if, for all $i\in\mathcal{I}$:
\[\langle \nabla_{i} J_i(x_i^*,x_{-i}^*),x_i-x_i^*\rangle\ge 0, \quad  \forall x_i \in X_i.\]
Using the Euclidean projection property, for an arbitrary scalar $\a>0$, the preceding relation is equivalent to \cite{FacchineiPang}:
\beqn
\label{eq-agent-fixed-point}
x_i^*=\Pi_{X_i}[x_i^*-\a \nabla_{i} J_i(x_i^*,x_{-i}^*)], \quad \forall i\in\mathcal{I}.
\eeqn

We define the game mapping as follows:
\begin{definition}\label{def:map_monotone}
The game mapping $F(x)\!\!:\!X\!\!\to\!\re^N$ is defined as
\begin{align}\label{eq:gamemapping}
F(x)\triangleq\left[\nabla_1 J_1(x_1,x_{-1}), \ldots, \nabla_N J_N(x_N,x_{-N})\right]^T.
 \end{align}
\end{definition}
\begin{lemma}\label{lem:map_monotone}
The game mapping is strongly monotone on $X$ with the constant $\mu=2\min_{i\in\mathcal{I}}Q_i+2\gamma >0$. Moreover, each function $J_i(x_i,x_{-i})$, $i\in\mathcal{I}$, is strongly convex on $\re$ for every $x_{-i}\in \re^{N-1}$ with the constant $\mu$.
\end{lemma}

As a result of Lemma~\ref{lem:map_monotone}, we now demonstrate the existence and uniqueness of the NE for the game $\Gamma$. For this purpose, we recall the results connecting NE and solutions of VIs \cite{FacchineiPang}.

\begin{definition}
For a set $K\subseteq \re^d$ and a mapping $g:K\to\re^d$, the variational inequality (VI) problem $VI(K,g)$ is to determine a vector $q^* \in K$ such that
\[\la g(q^*),q-q^*\ra \ge 0, \text{ for all } q \in K.\]
The set of solutions to $VI(K,g)$ is denoted by $SOL(K,g)$.
\end{definition}

The theorem below is a well-known result that establishes the connection between NE in games and solutions of a VI.

\begin{theorem}[Proposition 1.4.2 of \cite{FacchineiPang}]
\label{theo:NEVI}
Consider a networked Nash game $\Gamma$. Suppose that the action sets of the MEDs $\{X_i\}$ are closed and convex, the cost functions $\{J_{i}\}$ are continuously differentiable and convex in $x_{i}$ for every fixed $x_{-i}$ from $X_{-i}$. A vector $x^*\!\in\! X$ is a NE for the game $\Gamma$ if and only if $x^*\!\in\! SOL(X,F)$, where $F$ is the game mapping defined by \eqref{eq:gamemapping}.
\end{theorem}

\noindent
The next result holds for VIs with strong monotone mappings.

\begin{theorem}[Proposition 2.3.3 of \cite{FacchineiPang}]
\label{theo:uniqueNEVI}
Given the $VI(K,g)$, suppose that $K$ is a closed, convex set and the mapping $g$ is continuous and strongly monotone. Then, the solution set $SOL(Q,g)$ is nonempty and is a singleton.
\end{theorem}

Taking into account Lemma~\ref{lem-convexGame}, Lemma~\ref{lem:map_monotone}, Theorem~\ref{theo:NEVI} and Theorem~\ref{theo:uniqueNEVI}, we obtain the following results.

\begin{theorem}
\label{theo:uniqueNE}
Consider the game $\Gamma=(\mathcal{I},\{J_i\},\{X_i\})$. There exists a unique NE in $\Gamma$. Moreover, the NE is the solution of $VI(X,F)$, where $F$ is the game mapping defined in \eqref{eq:gamemapping}.
\end{theorem}

Theorem~\ref{theo:uniqueNE} guarantees the existence and uniqueness of NE in the game $\Gamma=(\mathcal{I},\{J_i\},\{X_i\})$. However, the formulation of the NE based on the VI $VI(X,F)$ does not consider the distributed nature of the proposed problem. We present the Lipschitz continuity of the gradient of local objective function $J_i$ for all $i\in\mathcal{I}$. This property is crucial to ensure the convergence of the distributed algorithm to the NE, as demonstrated in \cite{nguyen2022distributed}.

\begin{lemma}\label{lem:lip}
Consider the game $\Gamma=(\mathcal{I},\{J_i\},\{X_i\})$, we have:
\begin{itemize}
\item[(a)]
 The mapping 
 $\nabla_i J_i(\cdot,x_{-i})$ is 
 Lipschitz continuous on $\re$ for every $x_{-i}\in \re^{N-1}$ with a uniform constant $L_1=2(\max_{i\in\mathcal{I}}Q_i+\gamma)>0$, for all $i \in \mathcal{I}$.
 \item[(b)] The mapping $\nabla_i J_i(x_i,\cdot)$ is Lipschitz continuous on
$\re^{N-1}$ for every $x_i\in \re$ with a uniform constant $L_2=\gamma\sqrt{N-1}>0$, for all $i \in \mathcal{I}$.
\end{itemize}  
\end{lemma}

\section{Privacy-Preserving Decentralized Algorithm}
\label{algorithm}
We begin this section by highlighting the significance of a fully distributed algorithm that ensures the privacy of the information of MEDs participating in a system. When MEDs have unrestricted access to the actions of others in the system, it is possible to compute an NE point using an iterative algorithm \cite{FacchineiPang}.
In particular, starting with some initial point $x_0^i\in X_i$, each MED $i$ updates its decision at time $k$ as follows: 
\begin{equation}\label{eq-basic-algo}
    x_{k+1}^i
=\Pi_{X_i}[ x_k^i - \a \nabla_i J_i(x_k^i,x_k^{-i})].
\end{equation}

While this algorithm converges to a NE of the game $\Gamma$, it requires every MED to have access to all other MEDs' decisions at all times, which can be impractical in real-world systems. Existing approaches also assume full information about competitors' cost functions, which may not be feasible. In the crowdsourcing mobile edge content caching problem under consideration, MEDs may not have access to all the information about other MEDs, or collecting such information may raise privacy concerns. In the following, we propose to employ a privacy-preserving decentralized algorithm to compute the NE, allowing MEDs to learn it in a decentralized manner, even with partial information exchange.

Consider the crowdsourcing mobile edge content caching and sharing problem presented in Section~\ref{system}, situated in a partial-information environment with privacy-concerned MEDs. In this scenario, each mobile MED's local objective function $J_i$ is private and is not shared with other MEDs or the platform. Moreover, each MED only possesses partial information about their opponents' actions through local communications. All traffic remains within the local network and is not routed through a centralized cloud. As a result, there is no centralized communication system between MEDs, and the information exchange occurs solely between neighboring devices. 

MEDs can communicate only locally, with a local subset of neighboring MEDs, through D2D communication technologies, such as WiFi-Direct\footnote{https://www.wi-fi.org/discover-wi-fi/wi-fi-direct, Access March 2023}, LTE-Direct\footnote{https://www.qualcomm.com/research/5g/4g, Access March 2023} and the interactions over time are constrained by a sequence of time-varying communication graphs. Specifically, when MEDs interact at time $k$, their interactions are constrained by an undirected graph $\G_k=(\mathcal{I},\E_k)$. The set of nodes is the MED set $\mathcal{I}$, and $\E_k$ is the set of undirected links. An unordered link $(i,j)$ indicates that MED $i$ can receive information from MED $j$, and vice versa. Given a graph $\G_k=(\mathcal{I},\E_k)$, we define the neighbor set for every MED $i$, as follows:
\[\cNik=\{j\in\mathcal{I} \mid (i,j)\in\E_k\}\cup \{i\}.\]
\begin{remark}\label{rem:self-loop}
The sets of neighbors $\cNik$ always include MED $i$, indicating that MED $i$ has knowledge of its own decision. Consequently, every node $i \in \mathcal{I}$ has a self-loop in each graph $\G_k=(\mathcal{I}, \E_k)$, for all $k \ge 0$.
\end{remark}
\noindent We assume the following regarding the communication graph:
\begin{assumption}\label{assum-Bconnected}
The time-varying undirected graph sequence $\{\G_k\}$ is $B$-connected. Specifically, there exists an integer $B\ge 1$ such that the graph with edge set 
$\E^B_k=\bigcup_{i=kB}^{(k+1)B-1}\E_i$
is connected for every $k\ge 0$.
\end{assumption}
Assumption~\ref{assum-Bconnected} ensures that after $B$-rounds of communication, there exists a path between any pair of MEDs in the system. This assumption guarantees that no MED is isolated from the rest of the system, which is essential for the efficient information exchange and operation of the system as well as the convergence of the algorithm in distributed systems. This assumption is easily satisfied as long as MEDs are within the coverage area of cellular or Wi-Fi networks. In practice, the crowdsourcing-based market relies heavily on the connectivity of MEDs to the Internet. In order for MEDs to participate in crowdsourcing-based content-sharing tasks, they must be connected to a network that allows them to communicate with the CP and other MEDs in the network.


Given the constraints on MEDs' access to others' actions in game $\Gamma$, we propose a fully-decentralized algorithm that respects the information access as dictated by the communication graphs $\G_k$. Specifically, each MED $i$ maintains a local variable $z^i_{k}=(z^{i1}_k,\ldots,z^{iN}_k)^\T\in\re^N$, where $z^{ij}_k$ is MED $i$'s estimate of the decision $x^j_k$ for MED $j\ne i$, while $z^{ii}_k=x_k^i$. Thus, $z_k^i$ comprises the decision $z^{ii}_k=x_k^i\in\re$ of MED $i$ and the estimate $z_k^{i,-i}\in\re^{N-1}$ of MED $i$ for the decisions of the other MEDs. At each time $k$, each MED $i$ sends its estimate $z_k^i$ and receives estimates $z^j_k$ from its neighbors $j\in\cNik$. MED $i$ then updates its own action $x_{k+1}^i$ and local estimate $z_{k+1}^i$ using the received information. It is worth noting that in the decentralized algorithm, instead of evaluating its gradient at actual decisions, as in $J_i(x_i,x_{-i})$ (see \eqref{eq-basic-algo}), each MED evaluates its gradient at local estimates. The method is outlined in Algorithm 1. 

\begin{table}[t!]
\vspace{-0.2cm}
\centering \normalsize
    \begin{tabular}{l}
    \hline
    \multicolumn{1}{c}{\textbf{Algorithm 1: DCrowdCache}}\\
    \hline
    MEDs are instructed to use step-size $\alpha$. \\ 
    Every MED $i\in\mathcal{I}$ initializes with arbitrary initial vectors\\ $z_0^{i,-i}\in\re^{N-1}$ and $x_0^i\in\re$.\\
    \textbf{for} $k=0,1,\ldots,$ every MED $i\in\mathcal{I}$ does the following:\!\!\\
    \emph{  } Receives $z^j_k$ and $|\cNjk|$ from neighbors $j\in\cNik$;\\
    \emph{  } Sends $z_k^i$ and $|\cNik|$ to neighbors $j\in\cNik$;\\
    \emph{ } Calculates the weights using Metropolis weights\\
    \emph{~~~~} $[W_k]_{ij}\!=\!\!\begin{cases}
1/ (1\!+\!\max\{|\cNik|,|\cNjk|\}, \!\!\!\!&\text{if } j\!\in\!\cNik\backslash \{i\},\\
0, &\text{if } j\!\not\in\!\cNik,\\
1-\sum_{\ell\in\cNik}[W_k]_{il}, &\text{if } j=i;
\end{cases}$\\
    \emph{ } Updates estimates $z_{k+1}^{i,-i}$ and the action $x_{k+1}^i$ by \\
    \emph{~~~~} $\displaystyle z_{k+1}^{i,-i}=\sum_{j=1}^N [W_k]_{ij}z^{j,-i}_k$,\\
    \emph{~~~~} $\displaystyle x_{k+1}^{i}\!=\!\Pi_{X_i}\!\Bigg[\sum_{j=1}^N [W_k]_{ij}[z_k^j]_i -\a\nabla_i J_i\Bigg(\!\sum_{j=1}^N [W_k]_{ij}z_k^j\!\Bigg)\!\Bigg]$;\\
    \textbf{end for}\\
    \hline
    \end{tabular}
    \vspace{-0.3cm}
\end{table}

The Metropolis weights introduced in \cite{xiao2006distributed} are employed as the weights $W_k$ in Algorithm 1. 
The Metropolis weights can be calculated in a distributed fashion with two rounds of communication between neighboring nodes, without requiring any global knowledge of the communication graph. The initial round entails each node calculating its degree by enumerating its instantaneous neighbors, while in the subsequent round, each node disseminates its degree information to all its neighbors. It is noteworthy that the proposed method obviates the requirement of global knowledge pertaining to the communication graph. Due to the undirected nature of $\G_k$, it can be readily observed that the weight matrix $W_k$ satisfies the following:
\begin{lemma}\label{rem:weightMatrix}  
For each $k\!\ge\!0$, the weight matrix $W_k$ is compatible with the graph $\G_k$, symmetric, and  doubly-stochastic, i.e., 
\[W_k\one=\one ~~\text{ and } ~~\one^\T W_k = \one^\T.\]
Furthermore, there exist a scalar $w\!>\!0$ such that $\min^{+}(W_k)\ge w$, for all $k\ge 0$.
\end{lemma}

\begin{remark}\label{rem:weightMatrix}  
In the absence of any additional knowledge on the graph structures, the minimal weight can be observed as $w=1/N$. If more information is available, such as the maximal in-degree $d$ of the nodes (MEDs) in any of the communication graphs, then $w=1/(d+1)$.
\end{remark}

Using the iterates $z^i_{k}=(z^{i1}_k,\ldots,z^{iN}_k)^\T$,  define matrices
\begin{equation}\label{eq-notat}
    \bz_{k}=[z^1_{k},\ldots,z^N_{k}]^\T, ~~ \hat{\bz}_{k}=\one(\hat{z}_{k})^\T,~~
    \bx^*=\one(x^*)^\T,
\end{equation}
where $\hat{z}_k\!=\frac{1}{N}\sum_{i=1}^N z_k^i$
and $x^*$ is an NE point of the game $\Gamma$. 

The following presents the primary convergence result. It asserts that the iterates generated by Algorithm~1 converge linearly to the NE of the game $\Gamma$.
\begin{theorem} \label{theo:convTheo}
Consider the game $\Gamma=(\mathcal{I},\{J_i\},\{X_i\})$ where $X_i=\left[0,C_i\right]\subseteq \mathbb{R}$ and the function $J_i(x_i,x_{-i})$ is defined in \eqref{eq:localObj}, for each MED $i\in\mathcal{I}$. 
For a sufficiently small constant step-size $\a>0$, the sequence $\{\bz_k,k\ge 0\}$ generated by Algorithm 1 converges to $\bx^*=\one (x^*)^\T$, where $x^*$ is the unique NE of the game $\Gamma$, with linear rate, i.e.,
\[\lim_{k\to\infty}\|\bz_k-\bx^*\|=0,\qquad \lim_{k\to\infty}\|x_k-x^*\|=0.\]
\end{theorem}

The technique known as the heavy-ball method, originally introduced in \cite{POLYAK19641}, has gained widespread use as a means of accelerating gradient-based methods to achieve faster convergence in distributed optimization \cite{Nguyen2023AccAB}, non-cooperative games \cite{Nguyen2023AccGame}, and aggregative games \cite{Fang2022}. In light of this, we propose integrating the heavy-ball momentum and consensus-based gradient method to seek a discrete-time NE for the game $\Gamma$ in a distributed manner. Our proposed method is presented in Algorithm 2, and its linear convergence for unconstrained games has been proven in \cite{Nguyen2023AccGame}. We demonstrate the linear convergence of Algorithm 2 through numerical examples.

\begin{table}[t!]
\vspace{-0.2cm}
\centering \normalsize
    \begin{tabular}{l}
    \hline
    \multicolumn{1}{c}{\textbf{Algorithm 2: DCrowdCache-m}}\\
    \hline
    MEDs are instructed to use step-size $\alpha$ and parameter $\beta$. \\ 
    Every MED $i\in\mathcal{I}$ initializes with arbitrary initial vectors\\ $z_0^{i,-i},z_{-1}^{i,-i}\in\re^{N-1}$ and $x_0^i, x_{-1}^i\in\re$.\\
    \textbf{for} $k=0,1,\ldots,$ every MED $i\in\mathcal{I}$ does the following:\!\!\\
    \emph{  } Receives $z^j_k$ and $|\cNjk|$ from neighbors $j\in\cNik$;\\
    \emph{  } Sends $z_k^i$ and $|\cNik|$ to neighbors $j\in\cNik$;\\
    \emph{ } Calculates the weights using Metropolis weights;\\
    \emph{ } Updates estimates $z_{k+1}^{i,-i}$ and the action $x_{k+1}^i$ by \\
    \emph{~~~~} $\displaystyle z_{k+1}^{i,-i}=\sum_{j=1}^N [W_k]_{ij}z^{j,-i}_k+\!\beta(z_k^{i,-i}\!-z_{k-1}^{i,-i})$,\\
    \emph{~~~~} $\displaystyle x_{k+1}^{i}\!=\!\Pi_{X_i}\!\Bigg[\sum_{j=1}^N [W_k]_{ij}[z_k^j]_i -\a\nabla_i J_i\Bigg(\!\sum_{j=1}^N [W_k]_{ij}z_k^j\!\Bigg)\!\Bigg]$\\
    \emph{~~~~}$\qquad ~~~~ +\beta_i(x_k^i-x^i_{k-1})$;\\
    \textbf{end for}\\
    \hline
    \end{tabular}
    \vspace{-0.3cm}\end{table}


\section{Numerical Results}
\label{results}

\subsection{Simulation Setting}
\label{simsetting}
\begin{figure}[t!]
\vspace{-0.5cm}
\centering
\subfigure[MEDs coverage areas]{
\includegraphics[width=0.2315\textwidth,height=0.15\textheight]{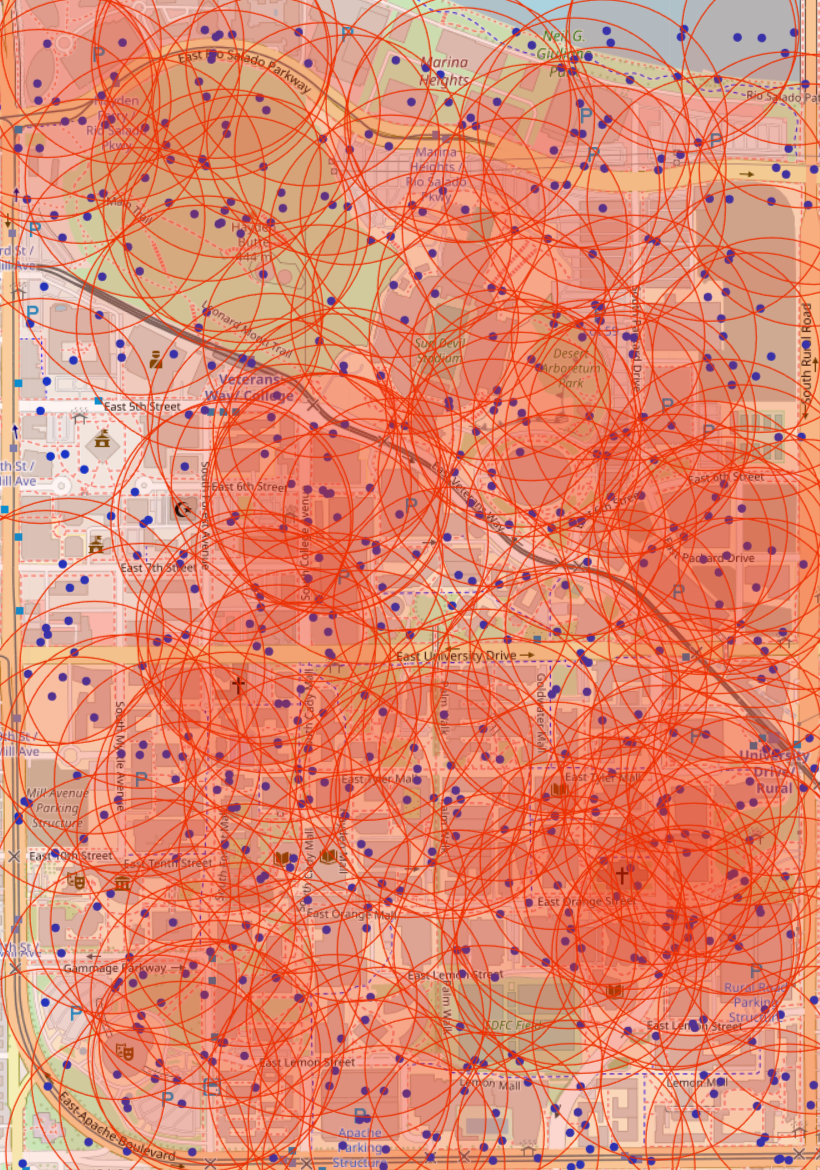}
\label{fig:ASU_Ranges}
	}   \hspace*{-1em} 
\subfigure[MEDs connections]{
\includegraphics[width=0.234\textwidth,height=0.15\textheight]{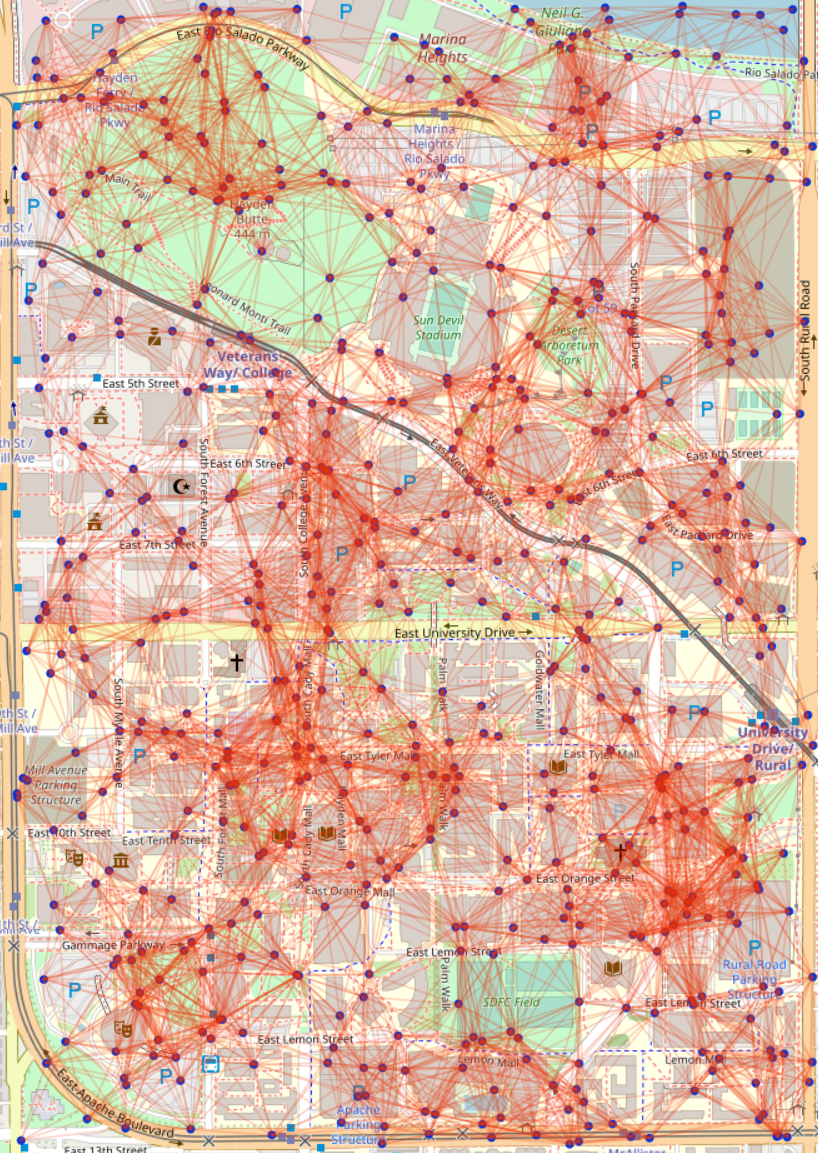}
\label{fig:ASU_Edges}}  \vspace{-0.2cm}
\caption{Mobile edge devices (MEDs) locations at ASU campus}
\label{fig:topology}
\vspace{-0.4cm}
\end{figure}

We consider a partial information scenario where there is no centralized communication system between MEDs. However, they may communicate with a local subset of neighbouring MEDs via undirected communication graph. We assume D2D communication among MEDs, such as WiFi-Direct or LTE-Direct. While the coverage for WiFi can vary depending on the device and environment, it typically has a range of up to $200$ meters in open spaces with no obstructions. In practical use, the range can be lower, especially in indoor environments where walls and other obstacles can interfere with the signal. We assume that the coverage range for each device is within the range of $[150, 200]$ meters.
By exploiting the EUA   dataset\footnote{https://github.com/swinedge/eua-dataset}, we generate initial locations of $N$ MEDs from the ASU Tempe campus, which includes longitude and latitude coordinates for each MED. 
In each iteration, the MEDs randomly move within the designated range, and their new locations and coverage ranges are used to create a communication graph among them. Figure~\ref{fig:topology} demonstrates the locations, coverage areas, and connections among MEDs around the ASU campus. We have made the raw experimental data publicly available, which includes information on the starting positions of MEDs, as well as their locations and connections over time \footnote{https://github.com/duongnguyen1601/CrowdCache-dataset}.

In our analysis, we begin with the \textbf{base case} scenario, which considers a small system comprising $N=2^9$ MEDs. We use the hourly price of the \textit{m5d.xlarge} Amazon EC2 instance\footnote{https://instances.vantage.sh, Access March 2023} as a reference point and normalize all parameters such that the maximum reward offered by the platform ($\bar{P}$) for each unit of storage is set to $1$. The quadratic cost of MED $i$ ($Q_i, \forall i$) is randomly generated from a uniform distribution $U[0.01,0.1]\$$ per hour, and the linear cost ($h_i, \forall i$) is generated from $U[0.05,0.15]\$$ per hour. We set the storage capacities $C_j$ randomly from the values $16$, $32$, $48$, and $64$ GB. The experiments are conducted in the MATLAB environment on a MacBook Air with an M2 core and 16 GB of RAM.

\subsection{Performance analysis}\label{subsec:Performance analysis}
It is well-known that under full information, the sequence of decision updates converges to the NE (see~\cite{FacchineiPang}). In this case, the NE $x^*$ can be found using 
\eqref{eq-basic-algo}. 
We evaluate the accuracy and efficiency of two proposed  privacy-preserving decentralized algorithms, DCrowdCache and DCrowdCache-m, and compare their performance with the state-of-the-art centralized algorithm, Cen-CrowdCache. Note that our proposed algorithms operate even under the partial information scenario, and the utility functions are private to each MED. However, Cen-CrowdCache is the algorithm when there is a centralized communication system that broadcasts information from all MEDs. 
We set the step-size $\alpha=20$ for all algorithms. The heavy-ball momentum parameter is $\beta=0.5$ for DCrowdCache-m(1) and $\beta=0.8$ for DCrowdCache-m(2). We compute the error between the iterations generated by the NE-seeking algorithm at iteration $k$ and the NE $x^*$ for each algorithm.

\begin{figure}[h!]
\vspace{-0.4cm}
\subfigure{
\includegraphics[width=0.253\textwidth]{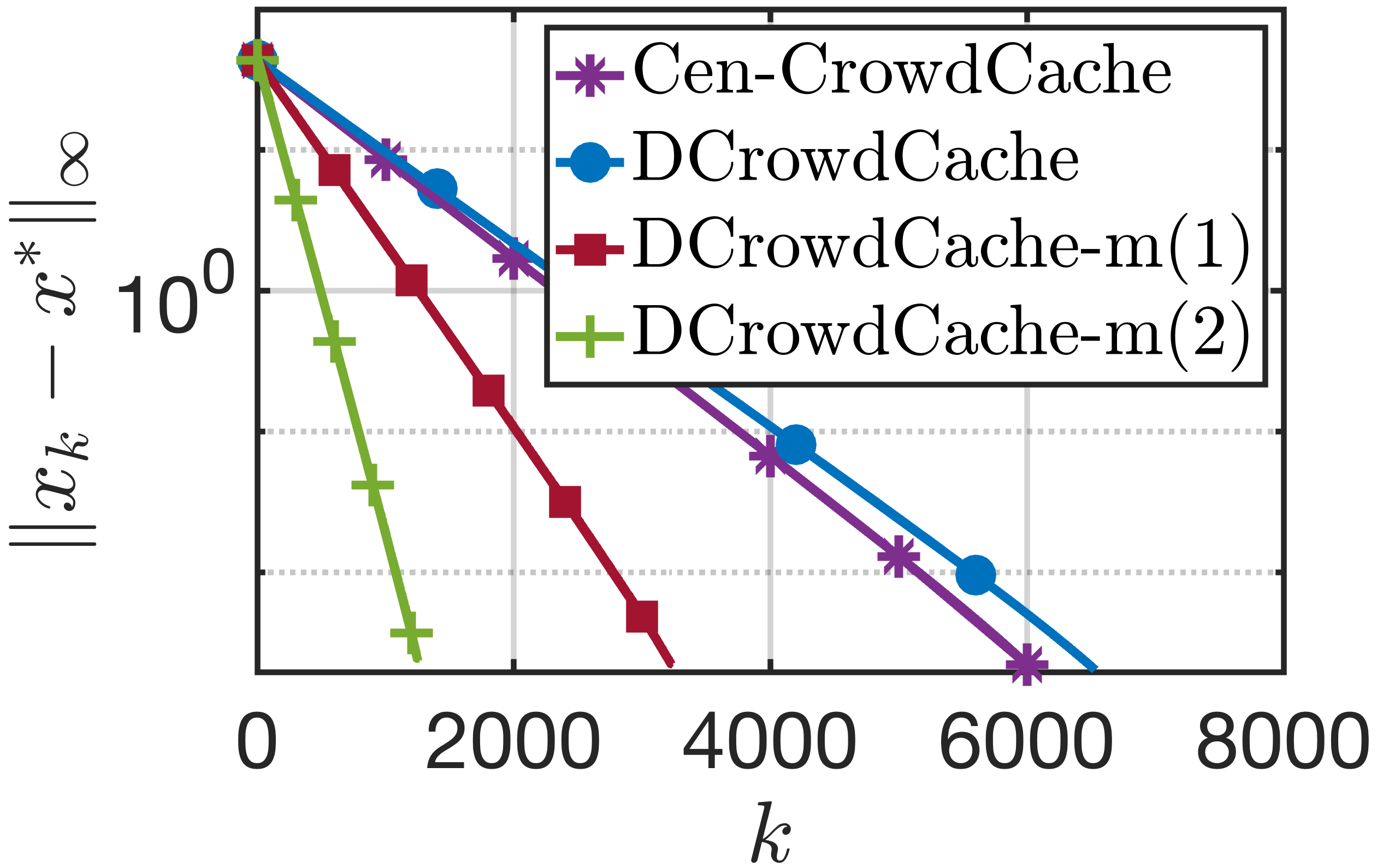}
\label{fig:convergence1}
	}    \hspace*{-1.2em} 
\subfigure{
\includegraphics[width=0.217\textwidth]{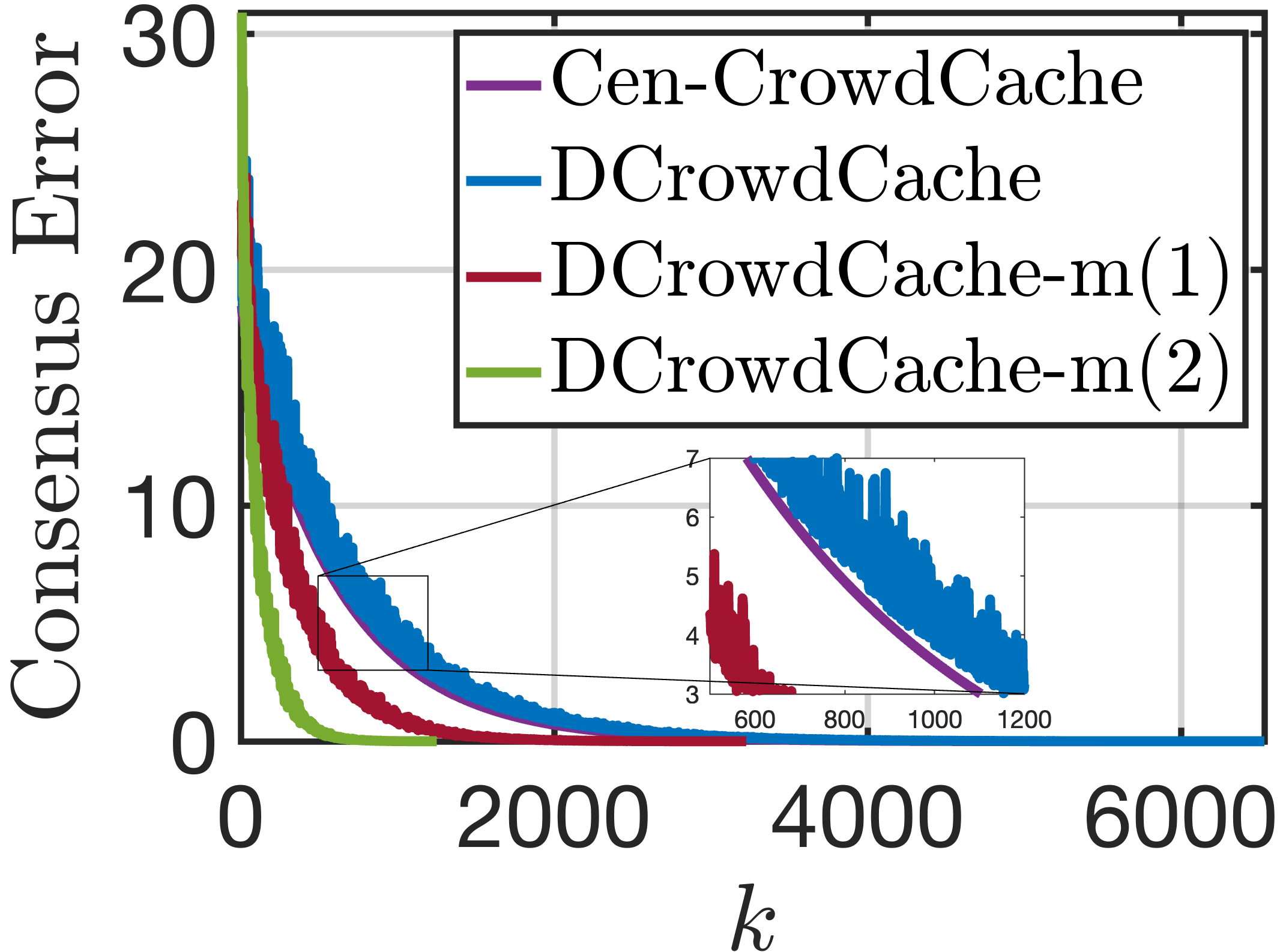}
\label{fig:convergence2}
	}  \vspace{-0.6cm}
	\caption{Convergence properties}
 \label{fig:convergence}
 \vspace{-0.4cm}
\end{figure}

Figure~\ref{fig:convergence} illustrates the convergence properties of all algorithms, indicating that the DCrowdCache-m algorithm converges faster than the DCrowdCache algorithm, and larger momentum leads to faster convergence. Notably, the proposed decentralized algorithms with privacy-preserving properties perform better than the centralized algorithm when heavy-ball momentum is used. When there are no momentum terms, the performance is still comparable which showcases the ability of the proposed fully distributed privacy-preserving algorithms to achieve effective coordination among MEDs.



\subsection{Sensitivity analysis}
We examine the impact of various design parameters on the system's performance. We use the term Proposed to refer to the proposed decentralized NE-seeking algorithms.
We compare three  algorithms: \textit{Proposed}, the heuristic scheme (\textit{Heuristic}), and the average scheme (\textit{Average}). The Heuristic algorithm uses $20\%$ of the maximum idle resources from MEDs, as long as the utility is positive. This percentage is hand-optimized. The Average algorithm uses $50\%$ of the maximum idle resources offered by MEDs, as long as the utility is positive. 

To quantify the effect of design parameters, we introduce scaling factors $\Gamma$, $\Lambda$ and $\Psi$ for the parameter $\gamma$, maximum unit price $\bar{P}$, quadratic cost  $Q_i$, for all $i\in\mathcal{I}$. The CP can dynamically adjust sensitivity parameter $\gamma$ to balance the trade-off between service availability and resource cost in MEC environment. For instance, when $\Gamma = 0.5$, the value of $\gamma$ used in the experiment is half of that used in the \textit{base case}. We conduct computations to derive and compare the average utility attained by individual MEDs, as well as the aggregated quantity of resources that are delegated from MEDs to the platform, for each of the considered algorithms.


\begin{figure}[h!]
\centering
\vspace{-0.4cm}
\subfigure{
\includegraphics[width=0.22\textwidth]{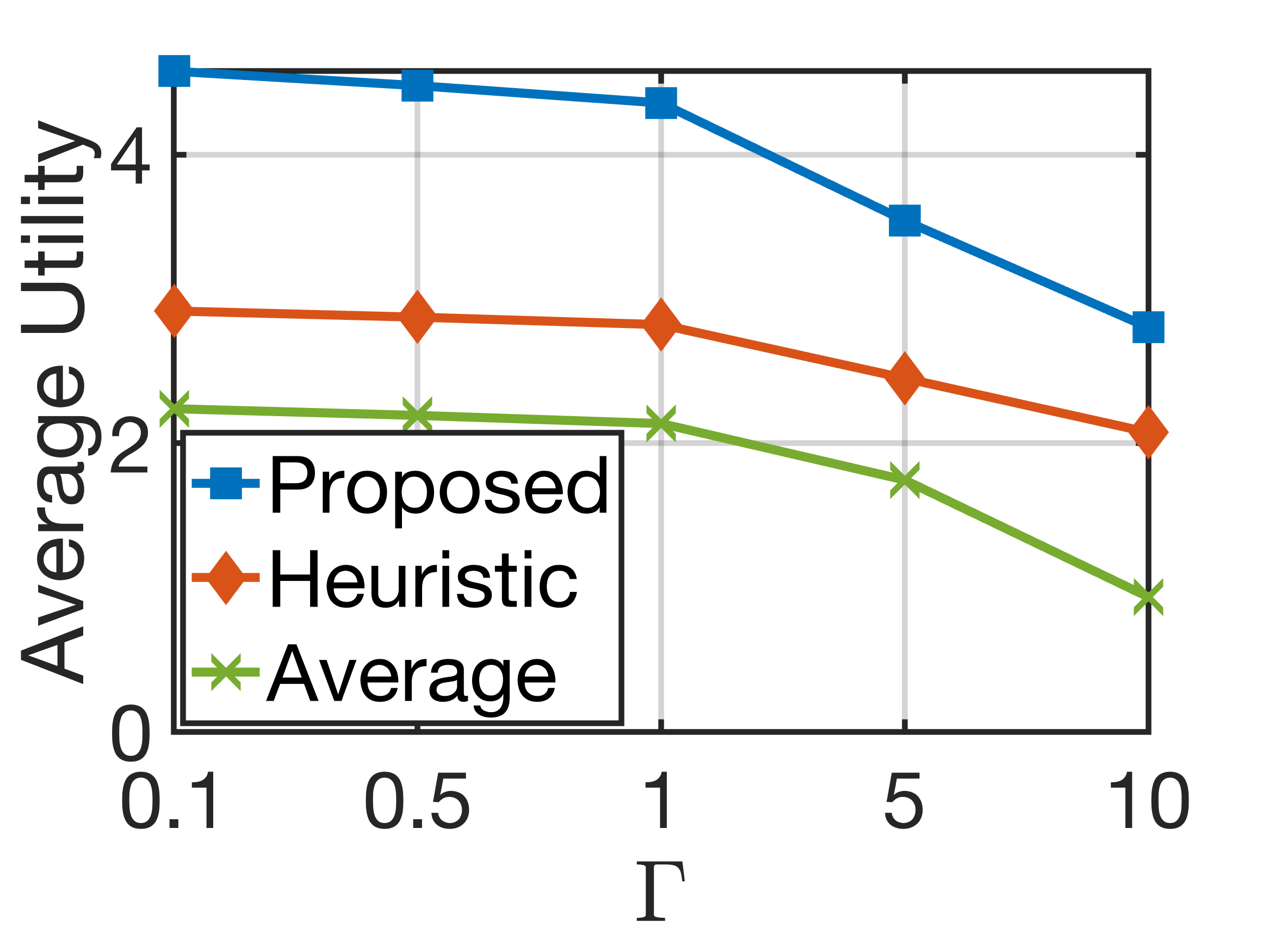}
\label{fig:UtilityGamma}
	}   \hspace*{-1em} 
\subfigure{
\includegraphics[width=0.22\textwidth]{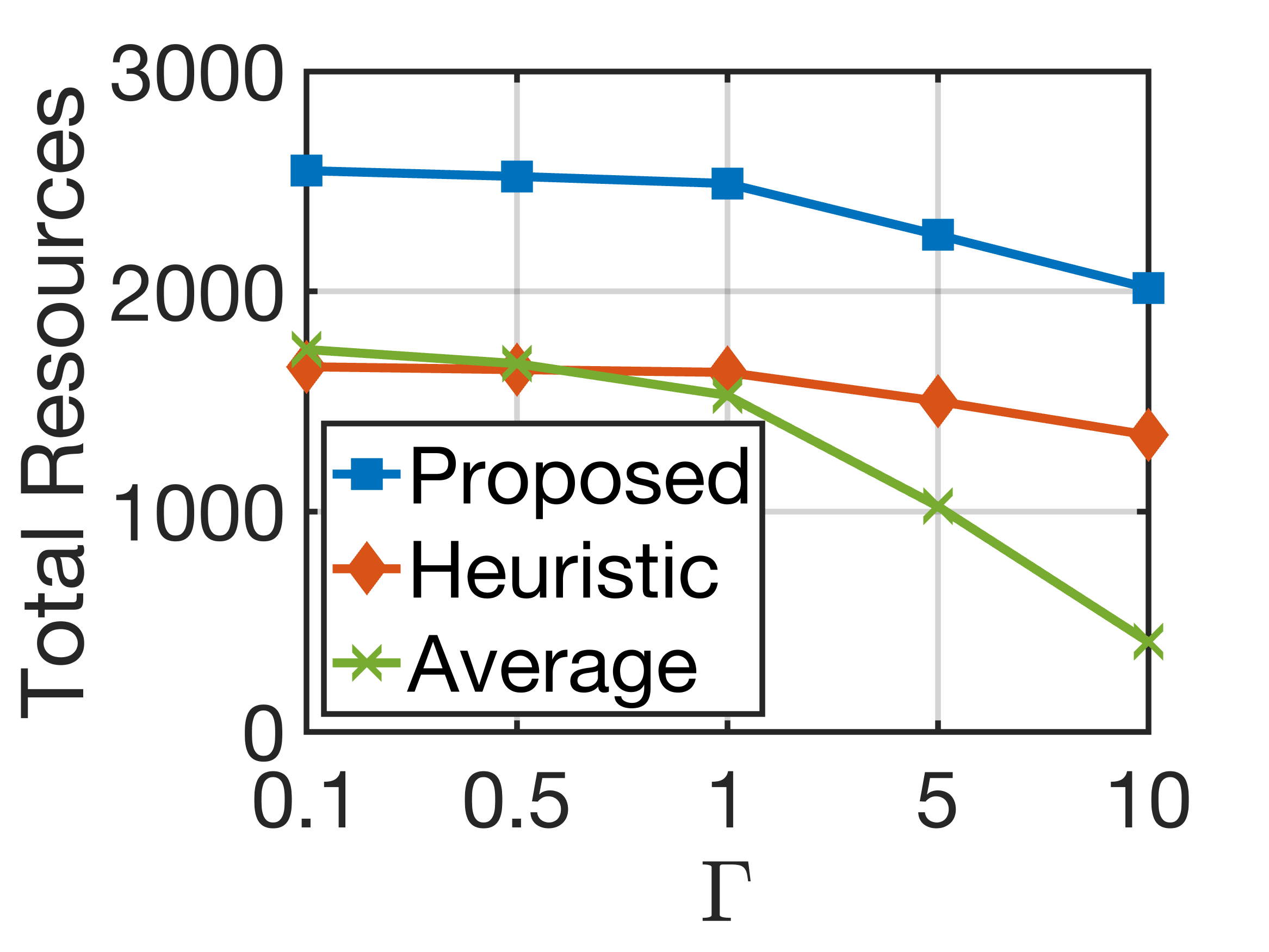}
\label{fig:ResourcesGamma}
	}  \vspace{-0.4cm}
\caption{Impacts of number of $\gamma$ }
\label{fig:gamma}
\vspace{-0.4cm}
\end{figure} 

Figure~\ref{fig:gamma} indicates that decreasing the value of $\gamma$ results in higher utility as expected since the unit price increases. When the CP reduces $\gamma$, it can motivate more MEDs to contribute their resources to the system, which in turn can enhance service availability and performance. In cases where the total demand for the content is already met by the available resources, increasing $\gamma$ can prompt MEDs to reduce their contributions, thus avoiding over-provisioning costs. 

\begin{figure}[h!]
\vspace{-0.4cm}
\centering
\subfigure{
\includegraphics[width=0.22\textwidth]{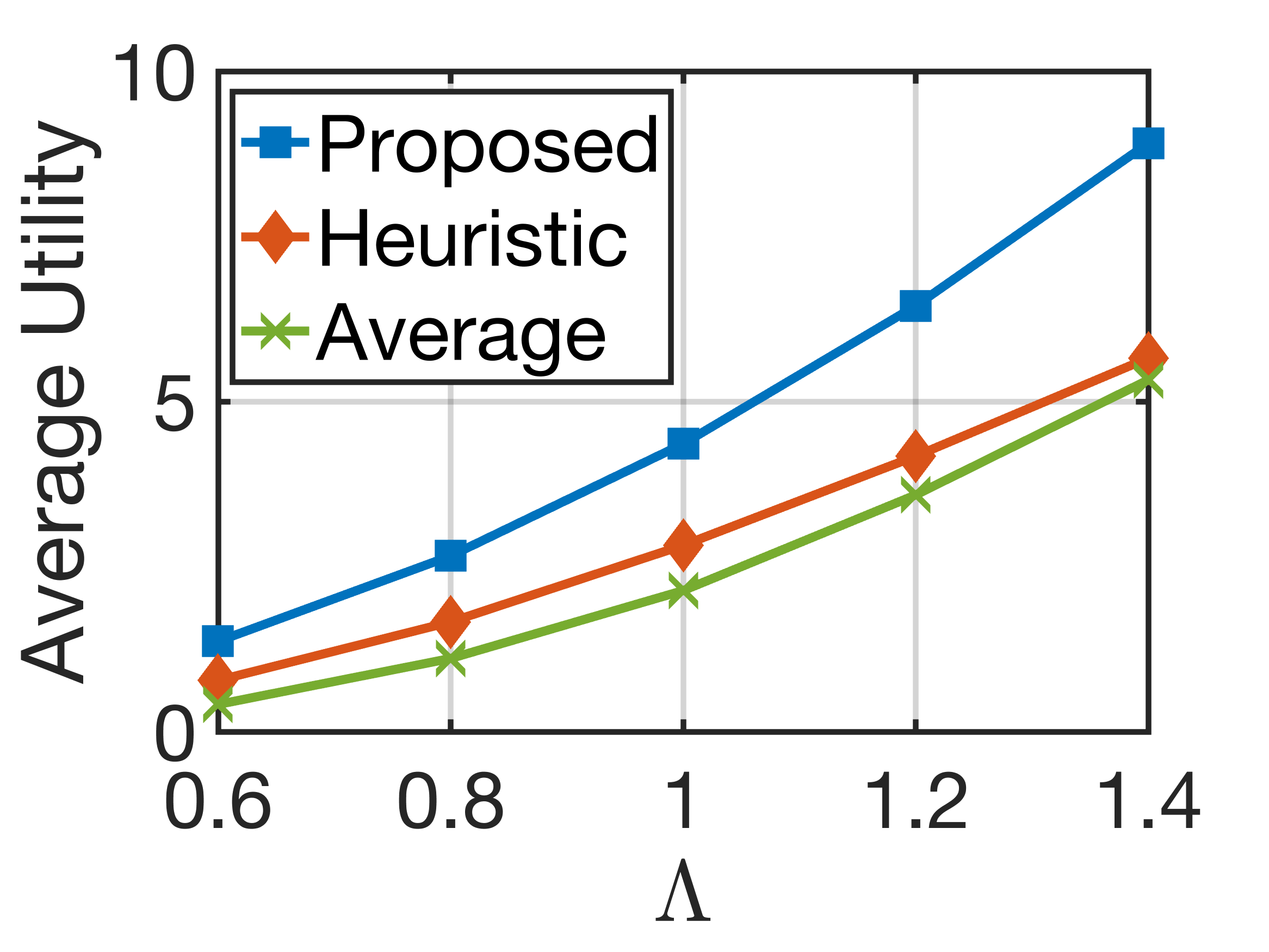}
\label{fig:UtilityCostRbar}
	}   \hspace*{-1em} 
\subfigure{
\includegraphics[width=0.22\textwidth]{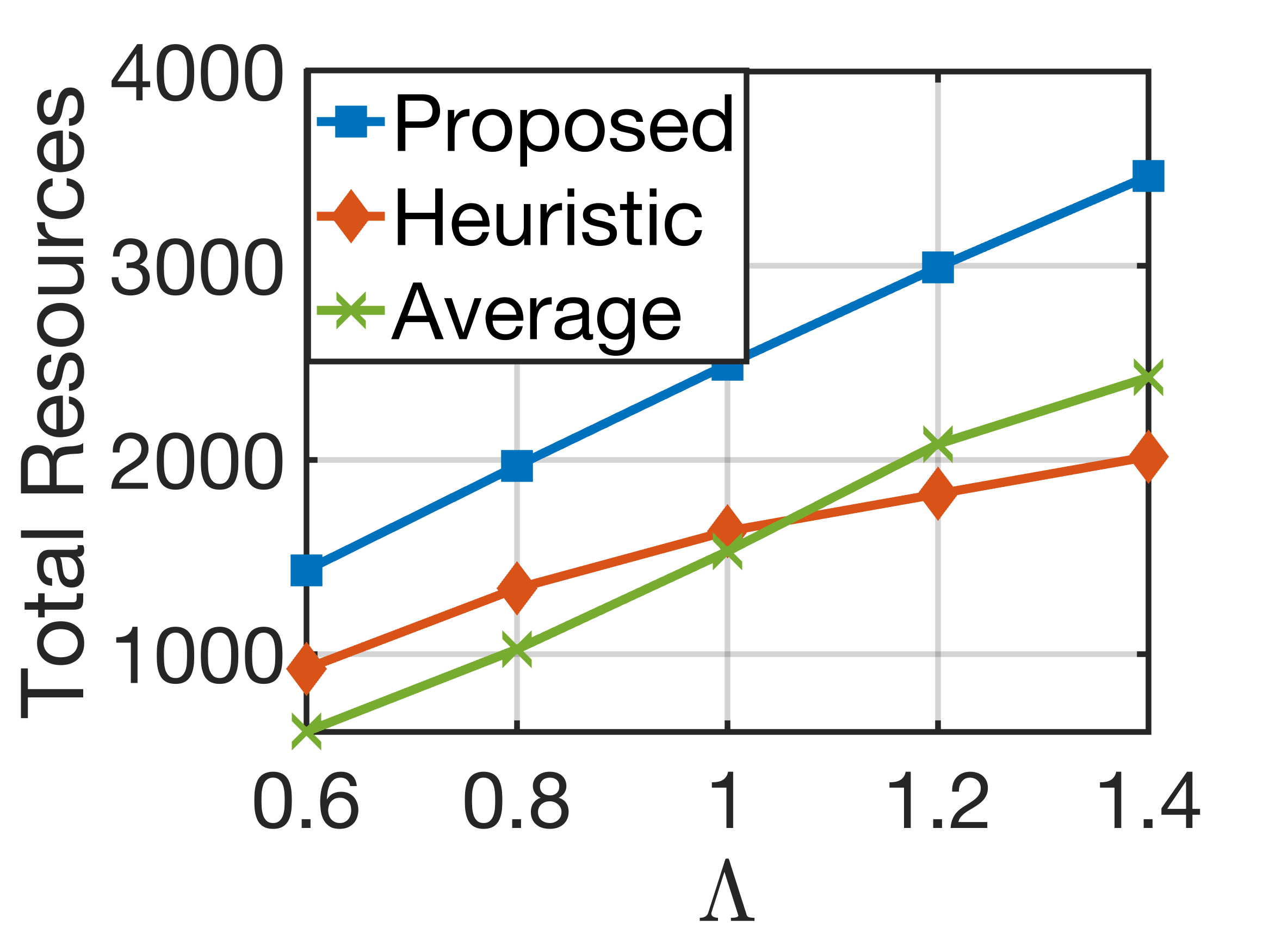}
\label{fig:ResourcesCostRbar}
	}  \vspace{-0.4cm}
\caption{Impacts of maximum unit price $\bar{P}$ }
\label{fig:Rbar}
\vspace{-0.4cm}
\end{figure} 

Determining the unit price for outsourced resources is a critical aspect of the platform. Several factors, including market demand, resource availability, competitive landscape, and quality, require thorough evaluation to set the optimal unit price that balances cost and profit objectives. The illustration in Figure~\ref{fig:Rbar} presents the effect of the maximum price $\bar{P}$ that the CP is willing to pay for one unit of resources. As expected, increasing $\bar{P}$ incentivizes the MED to supply more resources to the platform, resulting in a higher average utility for the MED, as shown in Figure~\ref{fig:UtilityCostRbar}. The platform benefits from this collaboration by accessing a larger pool of resources (see Figure~\ref{fig:ResourcesCostRbar}), leading to operational scalability, efficiency, and improved services to its users. 

\begin{figure}[h!]
\vspace{-0.4cm}
\centering
\subfigure{
\includegraphics[width=0.22\textwidth]{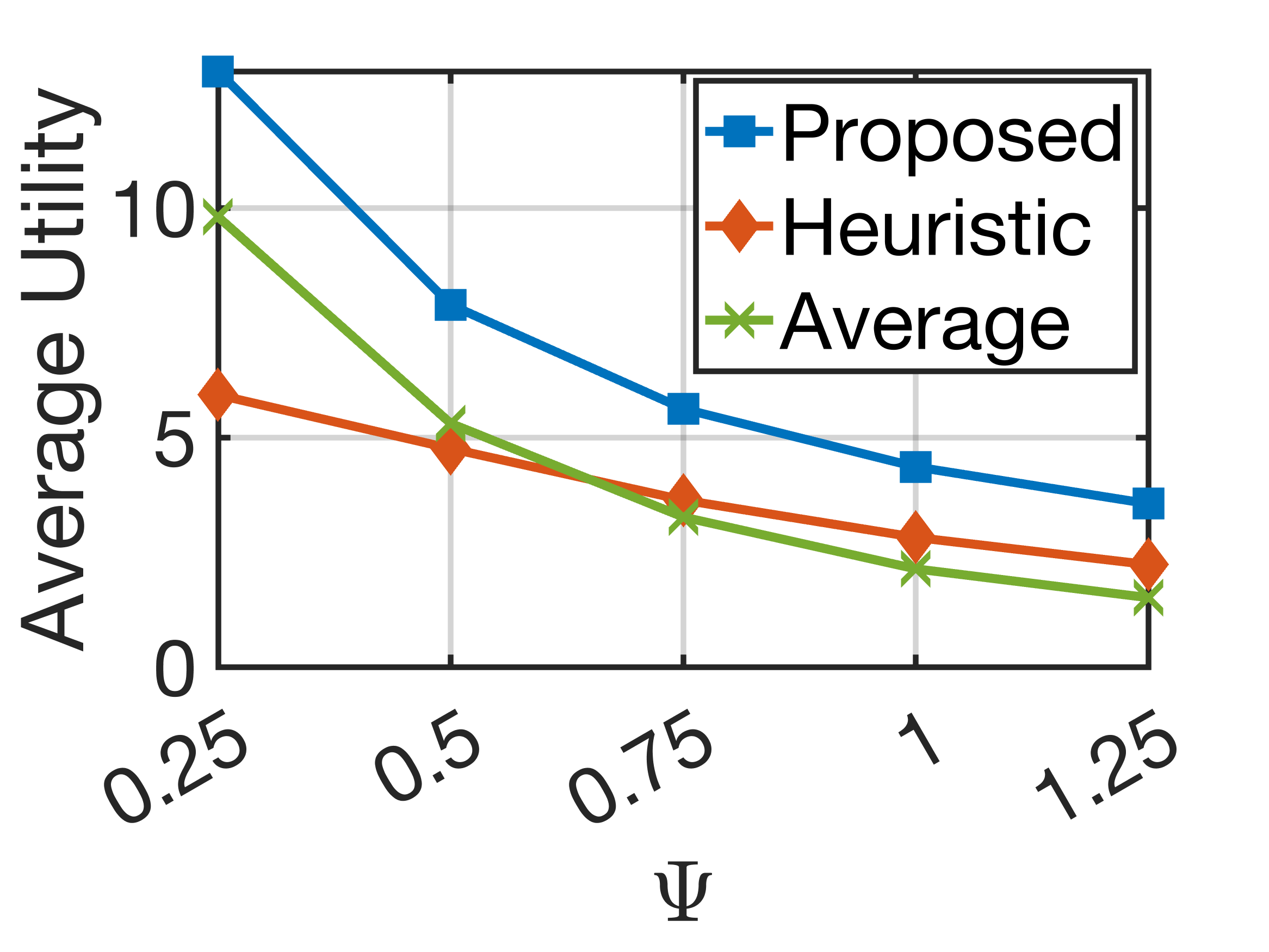}
\label{fig:UtilityCostQ}
	}   \hspace*{-1em} 
\subfigure{
\includegraphics[width=0.22\textwidth]{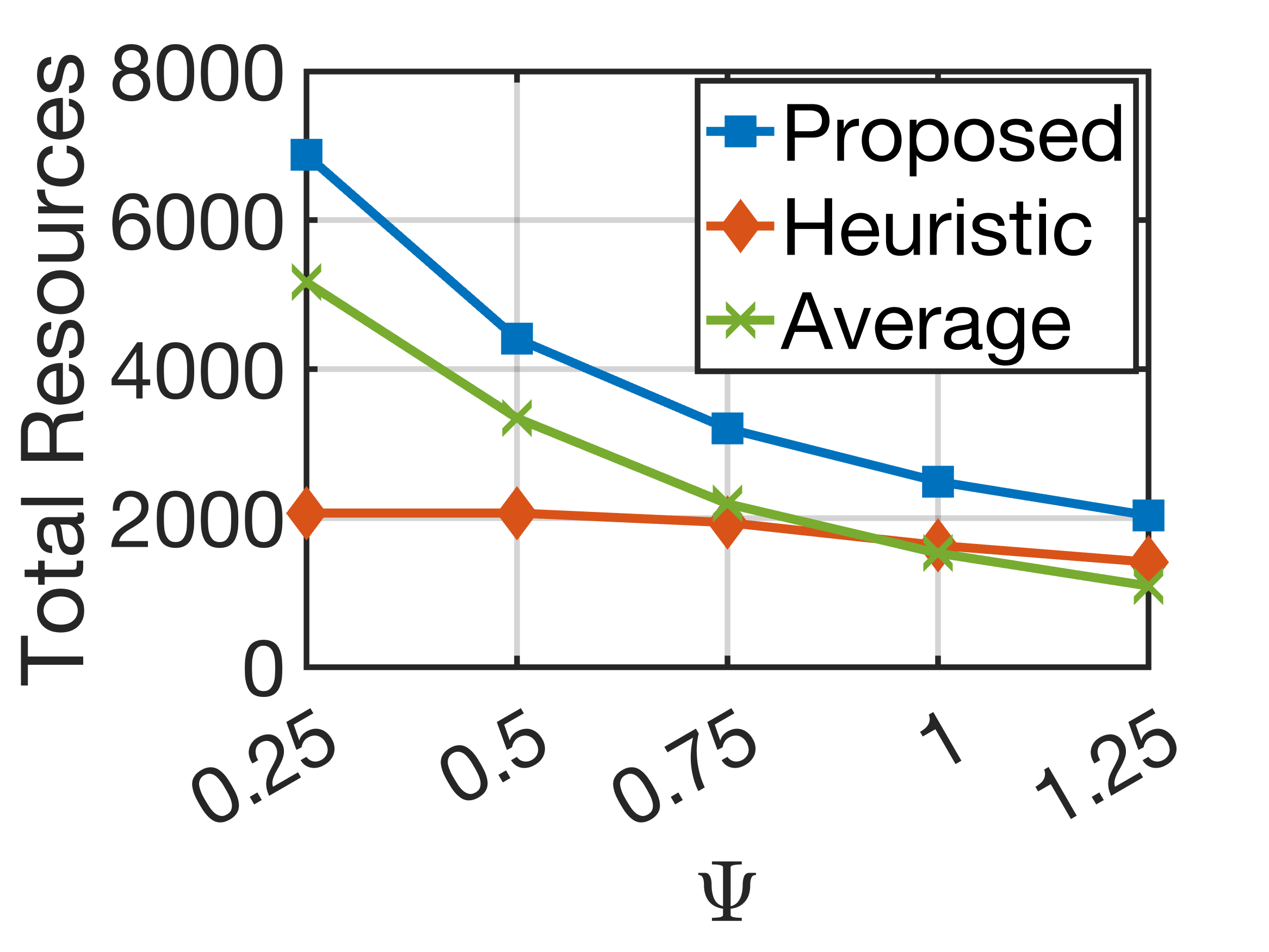}
\label{fig:ResourcesCostQ}
	}  \vspace{-0.4cm}
\caption{Impacts of $Q_i$, for all $i\in \mathcal{I}$ }
\label{fig:CostQ}
\vspace{-0.4cm}
\end{figure} 

Figure~\ref{fig:CostQ} illustrates the the impact of the quadratic cost coefficient ($Q_i$) associated with outsourcing additional resources. A higher quadratic cost coefficient results in a more significant increase in the cost of outsourcing additional resources. This, in turn, implies that the MED may reduce the number of resources it offers to outsource, leading to a decrease in the amount of idle resources available for outsourcing, as illustrated in Figure~\ref{fig:ResourcesCostQ}. Furthermore, the quadratic cost coefficient affects the MED's utility, which is a concave function of the number of resources produced. An increase in the quadratic cost coefficient leads to a more rapid decline in the marginal utility of outsourcing additional resources, as demonstrated in Figure~\ref{fig:UtilityCostQ}. 



\subsection{Time complexity analysis}
To analyze the time complexity, we conduct experiments with varying numbers of MEDs. We select different sizes of MEDs, including $N=2^8,2^9,2^{10},2^{11},$ and $2^{12}$, and measured the computation time required for each MED to execute the algorithm. The experimental results are presented in Table~\ref{Tab:Run_time}. As the table illustrates, the proposed system is highly efficient and can handle large-scale computations within a short time, particularly when a heavy-ball momentum is incorporated into the algorithm. The scalability of the proposed system with the problem size is also demonstrated. 

\begin{table}[ht!]
\centering
\begin{tabular}{|c|c|c|c|c|}
\hline
\multirow{2}{*}{$N$} & \multicolumn{2}{c|}{\textbf{DCrowdCache}}           & \multicolumn{2}{c|}{\textbf{DCrowdCache-m} ($\beta=0.5$)}    \\ \cline{2-5} 
&  \!\!\# Iterations\!\!    & \!\!Run time \!(s)\!\! & \!\! \# Iterations \!\!  & \!\!Run time \!(s)\!\! \\ \hline
$2^8$ & $2973$  &$0.035$ & $1474$ &$0.018$   \\ \hline
$2^9$ & $5791$  &$0.099$ & $2876$ &$0.053$    \\ \hline
$2^{10}$ & $10812$ &$0.437$  & $5407$ &$0.239$    \\ \hline
$2^{11}$ & $19951$ &$1.519$  & $9978$ &$0.858$   \\ \hline
$2^{12}$ & $37541$ & $8.482$ & $18734$ &   $4.464$  \\ \hline
\end{tabular}
\caption{Average performance over $1000$ simulations.}
\label{Tab:Run_time} 
\vspace{-0.5cm}
\end{table}

\section{Related work}

A considerable body of research has explored the topic of mobile edge content caching and sharing. For example, \cite{jiang20} formulates a crowdsourcing-based mobile edge caching problem as a Stackelberg game. Reference \cite{Chen18} considers a joint caching and resource allocation problem solved by a distributed ADMM-based algorithm. A computational offloading problem among MEDs is examined in \cite{chen14}, where the proposed decentralized game formulation allows MED users to make decisions locally.  \cite{hfeng21} develops a collaborative data caching and task offloading strategy in MEC. Additionally, \cite{Liu19} proposes an edge resource pooling framework and \cite{Lyun21} investigates content caching and user association in MEC. However, these studies assume complete information on competitors' actions, which is challenging to obtain in practical scenarios.



Privacy is a crucial issue in EC that has received increasing attention, but it has been overlooked in the existing literature. In \cite{Duong_TON}, authors develop a privacy-preserving distributed algorithm for computing an ME while allowing market participants to obfuscate their private information. Reference \cite{lbo21} suggests a distributed consensus mechanism to ensure edge data integrity and repair corrupted data replicas. In \cite{mpccache}, a privacy-preserving Multi-Party Cooperative Cache sharing framework is presented for multiple network operators. However, existing work typically assumes a time-invariant communication network among MEDs,  which may not hold in certain applications.

\section{Conclusion}
\label{conclusion}
The paper introduced a novel privacy-preserving framework to tackle the decentralized mobile edge content caching and sharing problem, where MEDs can share their cached content with neighbors. 
The framework was designed to operate over a time-varying communication graph, ensuring that users' privacy is preserved throughout the process. Specifically, we developed a decentralized gradient-based algorithm, called DCrowdCache, for NE computation. 
The algorithm achieves this by enabling users to exchange local information with their neighbors while performing a gradient step to maximize their utility. We also examined the convergence of the algorithm over a time-varying undirected communication network. Extensive numerical results demonstrated the efficacy of our proposed framework. 

\bibliographystyle{IEEEtran}
\bibliography{ref.bib}

\newpage 

\appendix
\subsection{Notations and Terminologies}\label{app:nota}
All vectors are viewed as column vectors unless stated otherwise and we consider a discrete time model where the time index is denoted by $k$. For every vector $u\in \mathbb{R}^N$, $u^\T$ is the transpose of $u$. We use $\langle \cdot,\cdot \rangle$ to denote the inner product, and $\|\cdot\|$ to denote the standard Euclidean norm. 
We write $\one$ to denote the vector with all entries equal to $1$. The dimensions of the vector $\one$ is to be understood from the context. 
The $i$-th entry of a vector $u$ is denoted by $u_i$, while it is denoted by $[u_k]_i$ for a time-varying vector $u_k$. Given a vector $v$, we  use $\min(v) $ and $\max(v)$ to denote the smallest and the largest entry of $v$, respectively, i.e., $\min(v)=\min_i v_i$ and $\max(v)=\max_i v_i$. A vector is said to be a stochastic vector if its entries are nonnegative and sum to $1$. 	

To denote the $ij$-th entry of a matrix $A$, we write $A_{ij}$, and we write $[A_k]_{ij}$ when the matrix is time-dependent. For any two matrices $A$ and $B$ of the same dimension, we write $A\le B$ to denote that $A_{ij}\le B_{ij}$, for all $i$ and $j$; in other words, the inequality $A\le B$ is to be interpreted component-wise. 
A matrix is said to be nonnegative if all its entries are nonnegative. For a nonnegative matrix $A$, we use ${\min}^{+}(A)$ to denote the smallest positive entry of $A$, i.e.,
denote ${\min}^{+}(A)=\min_{\{ij:A_{ij}>0\}} A_{ij}$. 

Given a vector $\pi\in\re^N$ with positive entries $\pi_1,\ldots,\pi_N$, the $\pi$-weighted inner product and $\pi$-weighted norm are defined, respectively, as follows:
\begin{center}
    $\la \bu,\bv\ra_{\pi}=\sum_{i=1}^N \pi_i\la u_i,v_i \ra \quad$and$\quad\|\bu\|_{\pi}=\sqrt{\sum_{i=1}^N \pi_i\|u_i\|^2},$
\end{center}
where $\bu:=[u_1^\T,\ldots,u_N^\T]^\T, \bv:=[v_1^\T,\ldots,v_N^\T]^\T \in \re^{N\times m}$, and $u_i,v_i\in\re^m$. When $\pi = \one$, we simply write $\la \bu,\bv\ra$ and $\|\bu\|$.

\subsection{Proof of Lemma~\ref{lem:map_monotone}} \label{app-map_monotoneProof}
\begin{proof}
We begin by computing the derivatives of $J_i$ with respect to $x_i$ for each MED $i\in \mathcal{I}$:
\[\nabla_i J_i(x_i,x_{-i}) = 2(Q_i+\gamma)x_i+\gamma \sum_{j \in \mathcal{I}\backslash\{i\}} x_j+h_i-\bar{P}.\]
For all $x=(x_i,x_{-i}),y=(y_i,y_{-i})\in \re^N$, we have
\begin{align*}
    &\sum_{i\in\mathcal{I}}\la x_i-y_i,\nabla_i J_i(x)-\nabla_i J_i(y)\ra\\
    =&\sum_{i\in\mathcal{I}} (2Q_i+\gamma)(x_i-y_i)^2+\gamma\Bigg(\sum_{j \in \mathcal{I}} (x_j-y_j)\Bigg)^2\\
    \ge& \left(2\min_{i\in\mathcal{I}}Q_i+2\gamma\right)\|x-y\|^2=\mu\|x-y\|^2,
\end{align*}
thus shows the strong monotonicity of the game mapping $F$. 

The above relation implies that for any $x=(x_i,x_{-i})\in \re^N$ and $y=(y_i,x_{-i})\in \re^N$, the following holds for all $i \in\mathcal{I}$:
\begin{align*}
    &\la x_i-y_i,\nabla_i J_i(x_i,x_{-i})-\nabla_i J_i(y_i,x_{-i})\ra\ge\mu\|x_i-y_i\|^2, 
\end{align*}
thus proves the strong convexity of function $J_i(x_i,x_{-i})$ on $\re$ for every $x_{-i}\in \re^{N-1}$.
\end{proof}

\subsection{Proof of Lemma~\ref{lem:lip}} \label{app-lipProof}
\begin{proof}
For any $x_{-i}\in\re^{N-1}$, we have for all $x_{i},y_{i}\in\re$:
\[|\nabla_i J_i(x_i,x_{-i})-\nabla_i J_i(y_i,x_{-i})| =2(Q_i+\gamma)|x_{i}-y_{i}|.\]
Similarly, for any $x_{i}\in\re$ and all $x_{-i},y_{-i}\in\re^{N-1}$, we have
\begin{align*}
&|\nabla_i J_i(x_i,x_{-i})-\nabla_i J_i(x_i,y_{-i})| \\
=~\gamma ~&\bigg|\sum_{j \in \mathcal{I}\backslash\{i\}} (x_j-y_j) \bigg|\le \gamma\sqrt{N-1}~\|x_{-i}-y_{-i}\|,
\end{align*}
which establish the desired statements in (a) and (b).
\end{proof}

\subsection{Basic Results} \label{app-BasicResults}
In our analysis of Algorithm~1, we use a mapping $\bF_\a(\cdot):\re^{N\times N}\to \re^{N\times N}$ to capture the updates for all MEDs $i\in\mathcal{N}$. Given a matrix $\bz\in\re^{N\times N}$, let $\bz_i$ be the vector in the $i$th row of 
$\bz$. The $i$th row of the matrix 
$\bF_\a(\bz)$ is defined by
\begin{equation}\label{eq-def-bfa}
[\bF_\a(\bz)]_{i:}=(0,\ldots, 0, \a(\nabla_i J_i(\bz_i))^\T, 0,\ldots,0 ).\end{equation}

\begin{lemma}[Lemma 5.6 of \cite{nguyen2022distributed}]\label{lemma-LipschitzbF}
Consider the mapping $\bF_\a(\cdot)$ defined by~\eqref{eq-def-bfa} and  
$L=\sqrt{L_1^2+L_2^2}$, we have
\begin{align*} 
    \!\!\|\bF_\a{(\bz)}-\bF_\a{(\by))}\|^2 \le L^2\a^2\|\bz-\by\|^2 \text{ for all } \bz,\by\in\re^{N\times N}.
\end{align*}
\end{lemma}

\begin{lemma}[\cite{nguyen2022distributed}, Corollary~5.2]\label{lem-normlincomb}
Consider a vector collection $\{u_i, \, i\in\mathcal{N}\}\subset\re^N$, and a scalar collection $\{\nu_i,\, i\in\mathcal{N}\}\subset\re$ of scalars such that $\sum_{i=1}^N \nu_i=1$. For all $u\in \re^N$, we have the following relation:
\begin{align*}
    \Bigg\|\!\sum_{i=1}^N \!\nu_i u_i - u \Bigg\|^2 \!\!\!= \!\sum_{i=1}^N \!\nu_i \|u_i-u\|^2 \!-\!\!\sum_{i=1}^N \nu_i \Bigg\|u_i \!-\! \Bigg(\sum_{j=1}^N \g_j u_j\Bigg)\!\Bigg \|^2\!\!\!.
\end{align*}
\end{lemma}

We have the following contraction property for a doubly-stochastic matrix $W$. The proof of this Lemma can be adapted from the proof of the contraction property of a row- or column-stochastic matrix in \cite{nguyen2022distributed}.
\begin{lemma}\label{lemma-lemma6PushPull}
Let $\G=(\mathcal{I},\E)$ be a connected undirected graph, and let $W$ be an $N\times N$ doubly-stochastic matrix that is compatible with the graph and has positive diagonal entries, i.e., $W_{ij}>0$ when $j=i$ and $(i,j)\in \E$, and $W_{ij}=0$ otherwise. Consider a collection of vectors $z_{1},\ldots,z_{N} \in \re^N$ and consider the vectors $r_i=\sum_{j=1}^N W_{ij}z_j$, for all $i\in\mathcal{N}$, and let $\hat{z}=\frac{1}{N}\sum_{i=1}^N z_i$, for all $u\!\in\! \re^N$, we have
\begin{align*}
    \sum_{i=1}^N\!\left\| r_i-u\right\|^2
    \!\!\le\! \sum_{j=1}^N\!\|z_j-u\|^2 \!\!-\!\frac{N\!\left({\min}^{+}(W)\!\right)^2}{\mathsf{D}(\G)\mathsf{K}(\G)}\!\sum_{j=1}^N\!\|z_j-\hat{z}\|^2\!.
\end{align*}
Here, $\mathsf{D}(\G)$ and $\mathsf{K}(\G)$ are the diameter and the maximal edge-utility of the graph $\G$, respectively.
\end{lemma}

\subsection{Proof of Theorem~\ref{theo:convTheo}} \label{app-convProof}
\begin{proof}
For simplicity, we provide the proof for $B=1$, i.e., every graph $\G_k$ is connected for all $k \ge 0$. The proof for $B>1$ follows similarly.

Under Theorem~\ref{theo:uniqueNE}, an NE point $x^*\in X$ exists and it is unique.
According to the notation in~\eqref{eq-notat},
we have that
\begin{align}\label{eq:ProjNorm1}
    \|\bz_{k+1}-\bx^*\|^2 \!=\!\sum_{i=1}^N\!\left(\|x_{k+1}^i-x_i^*\|^2 \!+ \|z_{k+1}^{i,-i}-x_{-i}^*\|^2\right)\!.\!\!
\end{align}
Using the definition of $x_{k+1}^i$ in Algorithm~1, the fixed point relation for the NE in~\eqref{eq-agent-fixed-point}, and
the non-expansiveness property of the projection, we obtain
\begin{align}\label{eq:ProjNorm2}
&\|x_{k+1}^i-x_i^*\|^2\nonumber \\
=&\Bigg\|\Pi_{X_i}\Bigg[\sum_{j=1}^N [W_k]_{ij}[z_k^j]_i -\a\nabla_i J_i\Bigg(\sum_{j=1}^N [W_k]_{ij}z_k^j\Bigg)\Bigg]\nonumber \\
&-\Pi_{X_i}[x_i^*-\a_i \nabla_{i} J_i(x^*)]\Bigg\|^2\nonumber \\
\le & \Bigg\|\sum_{j=1}^N [W_k]_{ij}[z_k^j]_i -\a\nabla_i J_i\Bigg(\sum_{j=1}^N [W_k]_{ij}z_k^j\Bigg)\nonumber \\
&-(x_i^*-\a_i \nabla_i J_i(x^*))\Bigg\|^2.
\end{align}

Combining \eqref{eq:ProjNorm1} and \eqref{eq:ProjNorm2}, we obtain
\begin{align}\label{eq-piNormEstFromWeightedAvgk}
    \|\bz_{k+1}-\bx^*\|^2 \le& \|(W_k\bz_k-\bx^*)-(\bF_\a{(W_k\bz_k)}-\bF_\a{(\bx^*)})\|^2\nonumber\\
    =& \|W_k\bz_k-\bx^*\|^2+\|\bF_\a{(W_k\bz_k)}-\bF_\a{(\bx^*)}\|^2\nonumber\\
    -&2\la W_k\bz_k-\bx^*,\bF_\a{(W_k\bz_k)}-\bF_\a{(\bx^*)} \ra ,
\end{align}
where $\bF_\a(\cdot)$ is as defined in~\eqref{eq-def-bfa}. We
apply Lemma~\ref{lemma-LipschitzbF} to bound the second term in the preceding relation, as follows:
\begin{align*} 
\|\bF_\a{(W_k\bz_k)}-\bF_\a{(\bx^*)}\|^2 \le L^2\a^2\|W_k\bz_k-\bx^*\|^2,
\end{align*}
where $L=\sqrt{L_1^2+L_2^2}$. Therefore,
\begin{align}
    \|\bz_{k+1}-\bx^*&\|^2
    \le (1+L^2\a^2) \|W_k\bz_k-\bx^*\|^2\cr
    &\ -2\la W_k\bz_k-\bx^*,\bF_\a{(W_k\bz_k)}-\bF_\a{(\bx^*)} \ra .\label{eq:piNormF}
\end{align}
To estimate the inner product in \eqref{eq:piNormF}, we write 
\begin{align} 
    &\la W_k\bz_k-\bx^*,\bF_\a{(W_k\bz_k)}-\bF_\a{(\bx^*)} \ra \nonumber\\
    =& \la W_k\bz_k-\bx^*,\bF_\a{(W_k\bz_k)}-\bF_\a{(\hat{\bz}_{k})} \ra \nonumber \\
    &+\la W_k\bz_k-\hat{\bz}_{k},\bF_\a{(\hat{\bz}_{k})}-\bF_\a{(\bx^*)} \ra \nonumber\\
    &+\la \hat{\bz}_{k}-\bx^*,\bF_\a{(\hat{\bz}_{k})}-\bF_\a{(\bx^*)} \ra . \label{eq:innerProd}
\end{align}
Applying the Cauchy–Schwarz inequality and Lemma~\ref{lemma-LipschitzbF}, we further obtain
\begin{align}\label{eq-est1}
    &\left|\la W_k\bz_k-\bx^*,\bF_\a{(W_k\bz_k)}-\bF_\a{(\hat{\bz}_{k})} \ra \right|\nonumber\\
    \le& \|W_k\bz_k-\bx^*\|\|\bF_\a{(W_k\bz_k)}-\bF_\a{(\hat{\bz}_{k})}\| \nonumber\\
    \le& \a L \|W_k\bz_k-\bx^*\|\|W_k\bz_k-\hat{\bz}_{k}\|.
\end{align}
Similarly,
\begin{align}\label{eq-est2}
    &\left|\la W_k\bz_k-\hat{\bz}_{k},\bF_\a{(\hat{\bz}_{k})}-\bF_\a{(\bx^*)} \ra \right| \nonumber\\
    \le & \|W_k\bz_k-\hat{\bz}_{k}\|\|\bF_\a{(\hat{\bz}_{k})}-\bF_\a{(\bx^*)}\| \nonumber\\
    \le &\a L \|W_k\bz_k-\hat{\bz}_{k}\|\|\hat{\bz}_{k}-\bx^*\|.
\end{align}

To estimate the last inner product in \eqref{eq:innerProd}, we write
\begin{align*} 
    &\la \hat{\bz}_{k}-\bx^*, \bF_\a{(\hat{\bz}_{k})}-\bF_\a{(\bx^*)} \ra \nonumber\\
    =& \a\sum_{i=1}^N \! \la\hat{z}_k^i-x_i^*, \nabla_i J_i(\hat{z}_k)- \nabla_i J_i(x^*)\ra \ge \a\mu\|\hat{z}_k-x^*\|^2,
\end{align*}
where the last inequality follows from Lemma~\ref{lem:map_monotone}. Thus,
\begin{align}\label{eq-est3}
    \la \hat{\bz}_{k}-\bx^*, \,\bF_\a{(\hat{\bz}_{k})}-\bF_\a{(\bx^*)} \ra \ge \frac{\a\mu}{N}\|\hat{\bz}_{k}-\bx^*\|^2.
 \end{align}
Upon substituting estimates~\eqref{eq:innerProd}--\eqref{eq-est3} into \eqref{eq:piNormF}, we obtain
\begin{align}\label{eq:EqPiTime}
    \|\bz_{k+1}\!\!-\!\bx^*\|^2 \!\!
    &\le\! \left(1\!+\!L^2\a^2\right)\!\!\|W_k\bz_k\!-\!\bx^*\|^2\!\!\!
    -\!\frac{2\mu\a}{N}\|\hat{\bz}_{k}\!-\!\bx^*\!\|^2\nonumber\\
    &+2\a L \|W_k\bz_k-\bx^*\|\|W_k\bz_k-\hat{\bz}_{k}\|\cr
    &+2\a L \|W_k\bz_k-\hat{\bz}_{k}\|\|\hat{\bz}_{k}-\bx^*\|.
\end{align}
The main idea of the rest of the proof is to determine the evolution relations for the quantity
$\|\bz_{k+1}-\bx^*\|$ in terms of $\|\bz_k-\hat{\bz}_{k}\|$ and $\|\hat{\bz}_{k}-\bx^*\|$.
Toward this end, we employ Lemma~\ref{lemma-lemma6PushPull} with $W=W_k$, $z_i=z_k^i\in\re^N$, $u=\hat{z}_k$ and notation 
$\hat{\bz}_{k}=\one (\hat{z}_k)^\T$ (see~\eqref{eq-notat}),
we obtain
\begin{equation}\label{eq-piNormEstFromWeightedAvg}
\|W_k\bz_k - \hat{\bz}_{k}\|^2
     \le c_k^2 \|\bz_k-\hat{\bz}_{k}\|^2,
\end{equation} 
where $c_k^2=\!\frac{Nw^2}{\mathsf{D}(\G_k)\mathsf{K}(\G_k)}\!\in \!(0,1)$. Similarly, applying Lemma~\ref{lemma-lemma6PushPull}  with $u=x^*$ and using notation 
$\bx^*=\one (x^*)^\T$ (see~\eqref{eq-notat}), yields
\begin{equation}\label{eq-piNormEstFromNash}
\|W_k\bz_k -\bx^*\|^2
     \le \|\bz_k - \bx^*\|^2 - (1-c_k^2) \|\bz_k-\hat{\bz}_{k}\|^2.
\end{equation}

By Lemma~\ref{lem-normlincomb}, with $\nu_i=\frac{1}{N}$, $u_i=z_k^i$, $u=x^*$, and noting that 
$\|\hat \bz_k- \bx^*\|^2=N\|\hat z_k- x^* \|^2$,
we find that 
\begin{equation}\label{eq-zaverandNash}
\|\bz_k-\bx^*\|^2
=\|\bz_k - \hat \bz_k\|^2
+\|\hat \bz_k- \bx^* \|^2.
\end{equation}
Plugging in the relation \eqref{eq-zaverandNash} for the first term in \eqref{eq-piNormEstFromNash}, we obtain the following 
\begin{equation}\label{eq-piNormEstFromNash1}
\|W_k\bz_k -\bx^*\|^2
     \le \|\hat \bz_k- \bx^* \|^2 +c_k^2 \|\bz_k-\hat{\bz}_{k}\|^2.
\end{equation}
Furthermore, by using $\sqrt{a+b}\le \sqrt{a}+\sqrt{b}$, which is valid for any two scalars $a,b\ge0$, the preceding relation implies
\begin{align}\label{eq-piNormEstFromNash2}
\|W_k\bz_k -\bx^*\| \le \|\hat \bz_k- \bx^* \|+ c_k \|\bz_k-\hat{\bz}_{k}\|.
\end{align}

Now, we are ready to finish the proof of the theorem. 
Substituting the preceding relations \eqref{eq-piNormEstFromWeightedAvg}, \eqref{eq-piNormEstFromNash1} and \eqref{eq-piNormEstFromNash2} back into \eqref{eq-piNormEstFromWeightedAvgk}, it follows that
\begin{align*}
    \|\bz_{k+1}-\bx^*\|^2
    &\le\left(1+L^2\a^2\right)(c^2\| \bz_k-\hat{\bz}_{k}\|^2+\|\hat{\bz}_{k}-\bx^*\|^2)\nonumber\\
    +&2\a L (c~\| \bz_k\!-\!\hat{\bz}_{k}\|\!\!+\!\|\hat{\bz}_{k}\!-\!\bx^*\|)c\|\bz_k\!-\!\hat{\bz}_{k}\|\nonumber\\
    +&2\a L c\|\bz_k-\hat{\bz}_{k}\|\|\hat{\bz}_{k}-\bx^*\| -\frac{2\mu\a}{N}\|\hat{\bz}_{k}-\bx^*\|^2 \nonumber\\
    =& \begin{bmatrix}
    \|\hat{\bz}_{k}-\bx^*\|\\ 
    \| \bz_k-\hat{\bz}_{k}\|
    \end{bmatrix}^\T \bar{Q}_\a \begin{bmatrix}
    \|\hat{\bz}_{k}-\bx^*\|\\ 
    \| \bz_k-\hat{\bz}_{k}\|
    \end{bmatrix},
    \end{align*} 
    where $c=\max_{k\ge 0}c_k \in (0,1)$ and,
\[\bar{Q}_\a=\begin{bmatrix}
1-\frac{2\mu\a}{N}+L^2\a^2 & 2cL\a \\
2cL\a  & (1+2L\a + L ^2\a^2)c^2
\end{bmatrix}.\]    
Hence,
\begin{align*}
\|\bz_{k+1}-\bx^*\|^2
    \le& \lambda_{\max}(\bar{Q}_\a) (\|\hat{\bz}_{k}-\bx^*\|^2+\| \bz_k-\hat{\bz}_{k}\|^2)\nonumber\\
    =&\lambda_{\max}(\bar{Q}_\a)\| \bz_k-\bx^*\|^2.\label{eq:convRela}
 \end{align*} 
where
$\lambda_{\max}(\bar{Q}_\a)$ is the largest eigenvalue of the matrix $\bar{Q}_\a$, and the last equality follows from \eqref{eq-zaverandNash}. If the step-sizes $\alpha$, $i\!\in\mathcal{I}$, are chosen such that $\lambda_{\max}(\bar{Q}_\a)\!<\!1$, 
then 
\[\lim_{k\to\infty}\|\bz_k-\bx^*\|=0,\qquad \lim_{k\to\infty}\|x_k-x^*\|=0,\]
with linear rate, which proves Theorem~\ref{theo:convTheo}.
\end{proof}

\subsection{Step-size Selection} \label{app-step-size}
The next remark provides the conditions to find the step-size $\a$, that guarantee the convergence of Algorithm 1, according to Theorem~\ref{theo:convTheo}.
\begin{remark}\label{remark-Sylvester}
$\lambda_{\max}(\bar{Q}_\a)<1$ if and only if the matrices $\bar{Q}_\a$ and $I-\bar{Q}_\a$ are positive definite. By Sylvester's criterion, the following inequalities should hold
\[\begin{cases}
[\bar{Q}_\a]_{1,1}>0,\\
\det(\bar{Q}_\a)>0,\\
[I-\bar{Q}_\a]_{1,1}>0,\\
\det(I-\bar{Q}_\a)>0,
\end{cases}\]
\end{remark}

We now illustrate that such a step-size can be found. By Sylvester's criterion in Remark~\ref{remark-Sylvester}, we have the following conditions for $\alpha$: 
\begin{align}
    L^2\alpha^2- 2\xi\alpha + 1&>0\label{eq:alpha1}\\
    c^2[L^4\alpha^4 + 2L^2(L-\xi)\a^3&\cr
    -2L(L+2\xi)\a^2+2(L-\xi)\a + 1]&>0\label{eq:alpha2}\\
   - L^2\alpha^2 + 2\xi\alpha&>0\label{eq:alpha3}\\
   \alpha[L^4c^2\alpha^3+ 2L^2(L - \xi)c^2\alpha^2 &\cr
    - (4L(L+\xi)c^2+L^2(1-c^2))\alpha + 2 (1-c^2)\xi]&>0\label{eq:alpha4}
\end{align}
with $L=\sqrt{L_1^2+L_2^2}$, $\xi=\frac{\mu}{N}$ and $c=\max_{k\ge 0}c_k \in (0,1)$. 

Note that form Lemma~\ref{lem:lip}, $\mu\le L_1$ and it follows that
$\xi<L$. As a consequence, the inequality~\eqref{eq:alpha1} holds for any $\alpha\in\re$.

Solving \eqref{eq:alpha3} leads to
\beqn \label{eq:solveAlpha1}
0<\alpha<\dfrac{2\xi}{L^2}.
\eeqn

Moreover, since the constant terms of the polynomials in \eqref{eq:alpha2} and \eqref{eq:alpha4} are positive, we can choose step-size $\alpha$ small enough that satisfies the two inequalities. Specifically, we have $L-\xi>0$, thus, \eqref{eq:alpha2} holds when 
$$L^4\alpha^4 -2L(L+2\xi)\a^2 + 1>0,$$
which gives
\begin{align} \label{eq:solveAlpha2}
    \a&<\dfrac{L(L+2\xi)-2L\sqrt{L\xi+\xi^2}}{L^4} \nonumber\\ \text{ or }~~\a&>\dfrac{L(L+2\xi)+2L\sqrt{L\xi+\xi^2}}{L^4}. 
\end{align}

The inequality in \eqref{eq:alpha4} holds if the quantity inside the square brackets is positive. By rearranging terms, we obtain:
\[c^2\Big[ L^4\a^2+2L^2(L-\xi)\a-4L(L+\xi) \Big]\a+(1-c^2)(2\xi-L^2\a)>0.\]
From \eqref{eq:solveAlpha1} we have $2\xi-L^2\a>0$ and since $c\in(0,1)$, we can either choose $\a$ very close to $0$ so that the term $(1-c^2)(2\xi-L^2\a)$ dominates or we can require the following inequality to hold $$L^4\a^2+2L^2(L-\xi)\a-4L(L+\xi)>0,$$
which is equivalent to have
\beqn \label{eq:solveAlpha3}
\alpha>\dfrac{-(L-\xi)+\sqrt{5L^2+2L\xi+\xi^2}}{L^2},
\eeqn
if it satisfies \eqref{eq:solveAlpha1} and \eqref{eq:solveAlpha2}.

\end{document}